\documentclass[aps,groupedaddress]{revtex4-1}
\usepackage{bm}
\usepackage{amsmath}
\usepackage{overpic}

\begin{document}


\title{Far-field theory for trajectories of magnetic ellipsoids \\in rectangular and circular channels}


\author{Daiki Matsunaga}
\email{daiki.matsunaga@physics.ox.ac.uk}
\author{Andreas Z\"ottl}
\author{Fanlong Meng}
\author{Ramin Golestanian}
\author{Julia M. Yeomans}
\affiliation{
 Rudolf Peierls Centre for Theoretical Physics, University of Oxford\\
 1 Keble Road, Oxford, OX1 3NP
}


\date{\today}

\begin{abstract}
We report a method to control the positions of ellipsoidal magnets in flowing channels of rectangular or circular cross section at low Reynolds number.
A static uniform magnetic field is used to pin the particle orientation, and the particles move with translational drift velocities resulting from hydrodynamic interactions with the channel walls which can be described using Blake's image tensor.
Building on his insights, we are able to present a far-field theory predicting the particle motion in rectangular channels, and validate the accuracy of the theory by comparing to numerical solutions using the boundary element method.
We find that, by changing the direction of the applied magnetic field, the motion can be controlled so that particles move either to a curved focusing region or to the channel walls.
We also use simulations to show that the particles are focused to a single line in a circular channel.
Our results suggest ways to focus and segregate magnetic particles in lab-on-a-chip devices.
\end{abstract}


\maketitle

\section{Introduction}
Understanding the dynamics of colloidal particles in microchannel flow is relevant to designing methods for guiding them in lab-on-a-chip devices \citep{Squires2005, Hejazian2015}.
At finite Reynolds number inertial effects, which lead to particle migration across stream lines, and to specific stable positions in the channel depending on particle size and Reynolds number \citep{DiCarlo2009}, can be used to focus the particles, and to sort them by size or shape.
At low Reynolds number, however, this mechanism ceases to be efficient and, in the absence of any external forces, both spherical and spheroidal particles simply move along streamlines without lateral migration \citep{Bretherton1962}.
In a shear flow spheres rotate with a constant angular velocity, whereas ellipsoidal particles follow Jeffrey orbits \citep{Jeffery1922} where their orientation unit vector performs periodic motion on the unit sphere.

Therefore, at low Reynolds number, external forces are needed to sort or focus colloidal particles. 
We recently showed that magnetic particles, subject to a uniform magnetic field, can be steered to an arbitrary position in channel flow \citep{Matsunaga2017}.
This occurs when the magnetic field suppresses the Jeffery rotation and pins the orientation of the particle.
Hydrodynamic interactions with the channel walls then lead to flow across streamlines, and the non-uniform hydrodynamic torque in the Poiseuille flow can be exploited to adjust the stable particle orientation, and hence its trajectory along the channel.

In reference \citep{Matsunaga2017} we considered the two-dimensional geometry of flow between two infinite, parallel plates. Here we extend our results to experimentally relevant three-dimensional situations, channels of rectangular and circular cross section. We compare an analytic far field theory and boundary element simulations to show that focusing is preserved in three dimensions. In a rectangular channel particles move either to a curved line or to the channel walls. In a cylindrical channel they are focused to a straight line.

\begin{figure}
    \centerline{\includegraphics[width=\columnwidth]{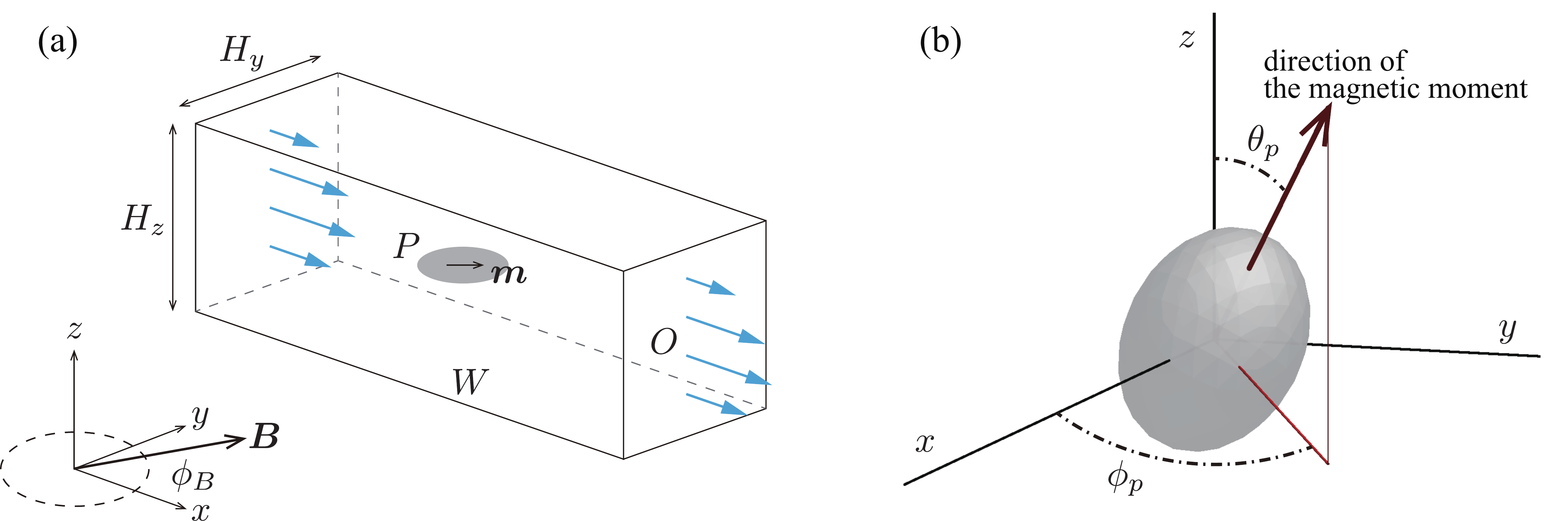}}
    \caption{(a) Schematic diagram of an ellipsoidal magnet flowing in an infinitely long rectangular channel. $\bm{m} = (m \sin \theta_p \cos \phi_p, m \sin \theta_p \sin \phi_p, m \cos \theta_p)$ is the magnetic moment. The flow is in the $+x$-direction and the magnetic field is in the ($x, y$)-plane: $\bm{B} = (B \cos \phi_B, B \sin \phi_B, 0)$.  $P$, $W$ and $O$ indicate the domains of the surface integral in Eq.~\eqref{eq:bie}. (b) Schematic diagram of the particle orientation ($\phi_p, \theta_p$). \label{fig:schematic}}
\end{figure}

\section{Problem statement and far-field theory}
We consider a permanent magnetic particle with prolate ellipsoidal shape of volume $4 \pi a^3/3$, suspended in an infinitely long rectangular channel with a width $H_y$ and a height $H_z$ as shown in Fig.~\ref{fig:schematic}(a).
The surrounding fluid is a Newtonian fluid of viscosity $\eta$ and density $\rho$.
The particle has magnetization $M$, and it is assumed to be neutrally buoyant for simplicity.
It has a single semi-axis of length $b_1 = a \alpha^{2/3}$ and two semi-axes with length $b_2 = a \alpha^{-1/3}$, where $\alpha$ is its aspect ratio, $\alpha = b_1/b_2 > 1$.
The particle has a magnetic moment parallel to its major axis given by $\bm{m} = (m \sin \theta_p \cos \phi_p, m \sin \theta_p \sin \phi_p, m \cos \theta_p)$ where $m = 4 \pi a^3 M/3$ is the magnetic moment, and $\theta_p, \phi_p$ are polar and azimuthal angles as shown in Fig.~\ref{fig:schematic}(b).
Hence the particle experiences a magnetic torque $\bm{T}_m = \bm{m} \times \bm{B}$ when a uniform external field $\bm{B}$ is applied to the whole domain.
We assume that $\bm{B}$ is oriented in ($x, y$)-plane as shown in Fig.~\ref{fig:schematic}(a), $\bm{B} = (B \cos \phi_B, B \sin \phi_B, 0)$, where $B$ is the strength and $\phi_B$ the orientation of the field which are both kept constant.

We introduce a non-dimensional parameter $\beta_w$ that describes the strength of the magnetic torque compared to the hydrodynamic torque as
\begin{equation}
    \beta_w = \frac{mB}{\eta a^3 \dot{\gamma}_w} = \frac{4 \pi MB}{3 \eta \dot{\gamma}_w}
\end{equation}
where $\dot{\gamma}_w$ is the characteristic wall shear rate inside a channel.
For example, when we assume that the particle magnetization $\mu_0 M = 10^{-3}$ $\rm{T}$ where $\mu_0 = 4\pi \times 10^{-7}$ $\rm{N/A^2}$ is the permeability of free space, particle size $a = 10^{-5}$ $\rm{m}$, water viscosity $\eta = 10^{-3}$ $\rm{Pa \cdot s}$, water density $\rho = 10^3$ $\rm{kg/m^3}$, shear rate $\dot{\gamma} = 10^2$ $\rm{s^{-1}}$ and magnetic field $B = 1.0^{-4} - 10^{-2}$ $\rm{T}$, the particle Reynolds number is $Re \approx 10^{-2}$ and the non-dimensional parameter $\beta \approx 10^{0} - 10^2$.

As presented in our previous work, which considered confinement between two infinite, planar walls leading to a flow translationally invariant along $z$ \citep{Matsunaga2017}, our strategy is to pin the particle orientation ($\phi_p, \theta_p$) by the external field $\bm{B}$, and to focus the particles to specific regions utilizing the drift velocities from the hydrodynamic interactions with the walls incorporated using image stresslets \citep{Blake1971}.
In this section, we introduce equations for the orientational dynamics and the translational velocities of an ellipsoidal magnet inside a rectangular channel.
To obtain the far-field expression, we assume that the particle size is much smaller than the channel size $a \ll H_y, H_z$, and that the particle is sufficiently far from the channel walls.
We then, in Section 3, compare the far-field theory to a full numerical simulation of the equations of motion.

\subsection{Flow field}
The velocity profile $v_x^\infty$ inside a rectangular channel with width $H_y$ ($0 \leq y \leq H_y$) and height $H_z$ ($-H_z/2 \leq z \leq +H_z/2$) is \citep{Mortensen2005}  
\begin{eqnarray}
    v_x^\infty (y, z) &=& V \sum_{n.odd}^{\infty} \frac{1}{n^3}
                   \left\{ 1 - \frac{\cosh(n \pi z/H_y)}{\cosh (n \pi H_z/(2 H_y))} \right\}
                   \sin \left( {n \pi y}/{H_y} \right), \label{eq:background} 
\end{eqnarray}
where $V$ is a parameter with the dimensions of velocity. The two shear components $\dot{\gamma}_{xy}$ and $\dot{\gamma}_{xz}$ follow as
\begin{eqnarray}
    \dot{\gamma}_{xy} (y, z) &=& \frac{\partial v_x}{\partial y} = V \sum_{n.odd}^{\infty} \frac{\pi}{n^2 H_y}
                   \left\{ 1 - \frac{\cosh(n \pi z/H_y)}{\cosh (n \pi H_z/(2 H_y))} \right\}
                   \cos \left( n\pi {y}/{H_y} \right), \\
    \dot{\gamma}_{xz} (y, z) &=& \frac{\partial v_x}{\partial z} = - V \sum_{n.odd}^{\infty} \frac{\pi}{n^2 H_y}
                   \frac{\sinh(n \pi z/H_y)}{\cosh (n \pi H_z/(2 H_y))}
                   \sin \left( n\pi{y}/{H_y} \right).
\end{eqnarray}
Defining a characteristic wall shear rate $\dot{\gamma}_w= \dot{\gamma}_{xy} (0,0)$ 
the velocity $V$ can be written 
\begin{equation}
    V = \frac{\dot{\gamma}_w}{\displaystyle \sum_{n.odd}^{\infty} \frac{\pi}{n^2 H_y} \left\{ 1 -  \frac{1}{\cosh (n \pi H_z/(2 H_y))} \right\}}.
\end{equation}

\subsection{Far field theory of the particle motion}
\subsubsection{Rotational motion}
The time evolution of the particle orientation ($\phi_p, \theta_p$) is governed by the background flow $\bm{v}^\infty$, the magnetic torque $\bm{T}_m$ and the particle-wall interactions.
When the particles are sufficiently far from the wall, the effect of the wall on the orientation of the particle is negligible \citep{Pozrikidis2005,Spagnolie2012} and the stable orientations ($\phi_p^*, \theta_p^*$) are obtained from the balance between the magnetic torque and the local hydrodynamic torque.
\cite{Almog1995} described the rotational motion of an ellipsoidal magnet subjected to a simple shear and a magnetic field.
Extending their equations to rotational motion under two shear components $\dot{\gamma}_{xy}$ and $\dot{\gamma}_{xz}$, which correspond to a flow field $\bm{v} = (y \dot{\gamma}_{xy} + z \dot{\gamma}_{xz}, 0, 0)$, gives
\begin{eqnarray}
    \frac{1}{\dot{\gamma}_w} \dot{\phi}_p   &=& \frac{\beta_w}{8 \pi} F(\alpha) \frac{\sin (\phi_B - \phi_p)}{\sin \theta_p}
                    -  \frac{\dot{\gamma}_{xy}}{2 \dot{\gamma}_w} (1 - J(\alpha) \cos 2\phi_p)
                    -  \frac{\dot{\gamma}_{xz}}{2 \dot{\gamma}_w} \frac{\sin \phi_p}{\tan \theta_p} (1 + J(\alpha)), \label{eq:d_phi_p}\\
    \frac{1}{\dot{\gamma}_w} \dot{\theta}_p &=& \frac{\beta_w}{8 \pi} F(\alpha) \cos \theta_p \cos(\phi_B - \phi_p)
                    +  \frac{\dot{\gamma}_{xy}}{4 \dot{\gamma}_w} J(\alpha) \sin 2 \theta_p \sin 2 \phi_p
                    +  \frac{\dot{\gamma}_{xz}}{2 \dot{\gamma}_w} \cos \phi_p (1 + J(\alpha) \cos 2\theta_p) \nonumber \label{eq:d_theta_p} \\
\end{eqnarray}
where $J(\alpha)$ and $F(\alpha)$ are shape functions \citep{Jeffery1922, Koenig1975},
\begin{eqnarray}
    J(\alpha) &=& \frac{\alpha^2 - 1}{\alpha^2 + 1}, \\ 
    F(\alpha) &=& \frac{3}{2 (\alpha^2 - \alpha^{-2})} \left\{ \frac{2\alpha - \alpha^{-1}}{\sqrt{\alpha^2 - 1}} \ln(\alpha + \sqrt{\alpha^2 - 1}) - 1 \right\}. 
\end{eqnarray}
Note that these equations assume that the particle size $a$ is much smaller than the channel dimensions $H_y$ and $H_z$
so that higher order contributions from the flow field, such as any change in the shear along the particle, can be ignored.

For large $\beta_w$, the particle reaches a single stable orientation ($\phi_p^*, \theta_p^*$) that satisfies $\dot{\phi}_p = 0$, $\dot{\theta}_p = 0$ \citep{Almog1995}. The stability conditions are
\begin{equation}
\left. \frac{1}{\dot{\gamma}_w} \frac{\partial \dot{\phi}_p}{\partial \phi_p} \right|_{\substack{\phi_p = \phi_p^*, \\ \theta_p = \theta_p^*}}<0, \quad\quad\quad\quad
\left. \frac{1}{\dot{\gamma}_w} \frac{\partial \dot{\theta}_p}{\partial \theta_p} \right|_{\substack{\phi_p = \phi_p^*, \\ \theta_p = \theta_p^*}} <0
\end{equation}
for each position ($y, z$) inside the channel.
By updating the angles ($\phi_p, \theta_p$) using equations \eqref{eq:d_phi_p} and \eqref{eq:d_theta_p} until they converge,
the stable orientation angle ($\phi_p^* (y, z), \theta_p^* (y, z)$) can easily be evaluated  numerically for any set of parameters \{$\alpha$, $\beta_w$, $\phi_B$, $H_y/a$, $H_z/a$\}.

\subsubsection{Translational motion}
Once the particles have reached the stable angles ($\phi_p^*, \theta_p^*$), they gain translational drift velocities from hydrodynamic interactions with the four surrounding walls which can be calculated using image systems \citep{Blake1971}.
In this rectangular channel, the leading order contribution to the particle translational velocities ($U_y, U_z$) comes from the stresslet $\bm{S}$ images \citep{Smart1991} as:
\begin{eqnarray}
    \frac{U_y}{\dot{\gamma}_w a} &=& - \frac{9}{64 \pi} \left\{ \frac{a^2}{y^2} - \frac{a^2}{(H_y - y)^2} \right\} \frac{S_{yy}}{\eta a^3 \dot{\gamma}_w}
                                     - \frac{3}{32 \pi} \left\{ \frac{a^2}{z^2} - \frac{a^2}{(H_z - z)^2} \right\} \frac{S_{yz}}{\eta a^3 \dot{\gamma}_w}, \label{eq:U_y} \\
    \frac{U_z}{\dot{\gamma}_w a} &=& - \frac{9}{64 \pi} \left\{ \frac{a^2}{z^2} - \frac{a^2}{(H_z - z)^2} \right\} \frac{S_{zz}}{\eta a^3 \dot{\gamma}_w}
                                     - \frac{3}{32 \pi} \left\{ \frac{a^2}{y^2} - \frac{a^2}{(H_y - y)^2} \right\} \frac{S_{yz}}{\eta a^3 \dot{\gamma}_w}. \label{eq:U_z}
\end{eqnarray}
When the particle is sufficiently far from the walls, the stresslets can be approximated by their values in free space, $\bm{S}^\infty$.
By extending the equations of \citet{KimBook}, the stresslet components in free space in the channel flow are
\begin{eqnarray}
    \frac{S_{yy}^\infty}{\eta a^3 \dot{\gamma}_{w}} (\alpha, \theta_p^*, \phi_p^*, \dot{\gamma}_{xy}, \dot{\gamma}_{xz})
           &=& \frac{5}{3} \pi \alpha^2 \frac{\dot{\gamma}_{xy}}{\dot{\gamma}_w} \sin 2 \phi_p^* \sin^2 \theta_p^* \{ C_1(\alpha) + C_2(\alpha) \sin^2 \phi_p^* \sin^2 \theta_p^* \} \nonumber \\
            && + \frac{5}{3} \pi \alpha^2 \frac{\dot{\gamma}_{xz}}{\dot{\gamma}_w} \cos \phi_p^* \sin 2 \theta_p^* \{ C_3(\alpha) + C_2 (\alpha) \sin^2 \phi_p^* \sin^2 \theta_p^* \} \nonumber \\
            && - 2 \pi \alpha^2 \frac{\dot{\gamma}_{xy}}{\dot{\gamma}_w} Y^H \sin 2 \phi_p^* \sin^2 \theta_p^*, \label{eq:Syy} \\ 
    \frac{S_{zz}^\infty}{\eta a^3 \dot{\gamma}_{w}} (\alpha, \theta_p^*, \phi_p^*, \dot{\gamma}_{xy}, \dot{\gamma}_{xz})
           &=& \frac{5}{3} \pi \alpha^2 \frac{\dot{\gamma}_{xy}}{\dot{\gamma}_w} \sin 2 \phi_p^* \sin^2 \theta_p^* \{ C_3 (\alpha) + C_2 (\alpha) \cos^2 \theta_p^* \} \nonumber \\
            && + \frac{5}{3} \pi \alpha^2 \frac{\dot{\gamma}_{xz}}{\dot{\gamma}_w} \cos \phi_p^* \sin 2 \theta_p^* \{ C_1 (\alpha) + C_2 (\alpha) \cos^2 \theta_p^*\} \nonumber \\
            && - 2 \pi \alpha^2 \frac{\dot{\gamma}_{xz}}{\dot{\gamma}_w} Y^H \cos \phi_p^* \sin 2\theta_p^*, \label{eq:Szz} \\
    \frac{S_{yz}^\infty}{\eta a^3 \dot{\gamma}_{w}} (\alpha, \theta_p^*, \phi_p^*, \dot{\gamma}_{xy}, \dot{\gamma}_{xz})
           &=& \frac{5}{3} \pi \alpha^2 \frac{\dot{\gamma}_{xy}}{\dot{\gamma}_w} \cos \phi_p^* \sin 2 \theta_p^* \{ C_4(\alpha) + C_2 (\alpha) \sin^2 \phi_p^* \sin^2 \theta_p^* \} \nonumber \\
            && + \frac{5}{3} \pi \alpha^2 \frac{\dot{\gamma}_{xz}}{\dot{\gamma}_{w}} \sin 2 \phi_p^* \sin^2 \theta_p^* \{ C_4 (\alpha) + C_2(\alpha) \cos^2 \theta_p^*\} \nonumber \\
            && - \pi \alpha^2 Y^H (\frac{\dot{\gamma}_{xy}}{\dot{\gamma}_w} \cos \phi_p^* \sin 2 \theta_p^* + \frac{\dot{\gamma}_{xz}}{\dot{\gamma}_{w}} \sin 2\phi_p^* \sin^2 \theta_p^*) 
\end{eqnarray}
where $C_1$, $C_2$, $C_3$, $C_4$ are shape functions
\begin{eqnarray}
    C_1 (\alpha) &=& - X^M + 2 Y^M - Z^M, \\
    C_2 (\alpha) &=& 3 X^M - 4 Y^M + Z^M, \\
    C_3 (\alpha) &=& - X^M + Z^M, \\
    C_4 (\alpha) &=& Y^M - Z^M
\end{eqnarray}
and $X^M$, $Y^M$, $Z^M$, $Y^H$ \citep{KimBook} depend on the eccentricity $e = \sqrt{1 - \alpha^{-2}}$ as
\begin{eqnarray}
    X^M (e) &=& \frac{8}{15} e^5 \frac{1}{(3 - e^2) L - 6e}, \\
    Y^M (e) &=& \frac{4}{5} e^5 \frac{2e(1 - 2e^2) - (1 - e^2) L}{(2e(2 e^2 - 3)+3(1 - e^2)L)(-2e + (1 + e^2)L)}, \\
    Z^M (e) &=& \frac{16}{5} e^5 \frac{1 - e^2}{3(1 - e^2)^2 L - 2e (3 - 5e^2)}, \\
    Y^H (e) &=& \frac{4}{3} e^5 \frac{1}{-2e +(1 + e^2)L}, \\
    L(e) &=& \ln \left( \frac{1 + e}{1 - e} \right).
\end{eqnarray}
Substituting the free-space expressions for the stresslets into equations \eqref{eq:U_y} and \eqref{eq:U_z} gives the far-field prediction of the particle translational velocity in the channel.
Note that the higher order reflections of the stresslets can be ignored by assuming sufficiently large channel sizes $H_y$ and $H_z$ compared to the particle size \citep{Mathijssen2016,Matsunaga2017}.

We will compare results from the approximate analytical theory to exact solutions of the equations of motion obtained using a boundary element approach in the next section once we have described the numerical scheme.

\section{Comparison of far-field theory and full numerical simulation}
In order to check the accuracy of the far-field theory, we compare results for the stable angle ($\phi_p^*$, $\theta_p^*$), the stresslet $\bm{S}$, and the drift velocity $(U_y, U_z)$ to a full simulation using the boundary element method.
In this section, we explain our numerical method, and present the comparison between the theory and the simulations.

\subsection{Numerical method}
\subsubsection{Governing equations}
When the Reynolds number is small and inertial effects are negligible, the pressure driven flow field $\bm{v}$ of a given point $\bm{x}$ inside the channel can be described using a boundary integral formulation \citep{Pozrikidis2005, Hu2012}:
\begin{equation}
    v_i(\bm{x}) = v_i^\infty (\bm{x}) - \frac{1}{8 \pi \eta} \left\{ \int_P G_{ij} (\bm{x}, \bm{y}) q_j (\bm{y}) dA + \int_W G_{ij} (\bm{x}, \bm{y}) q_j(\bm{y}) dA + \Delta P \int_O G_{ij} (\bm{x}, \bm{y}) n_j (\bm{y})dA \right\} \label{eq:bie}
\end{equation}
where
\begin{equation}
    G_{ij}(x, y) = \frac{\delta_{ij}}{|\bm{r}|} + \frac{r_i r_j}{|\bm{r}|^3}
\end{equation}
is the Green's function, $\bm{r} = \bm{x} - \bm{y}$ is the relative position vector, $\bm{v}^\infty$ is the background flow \eqref{eq:background}, $\bm{n}$ is the normal vector pointing into the channel, and $\bm{q}$ is the viscous traction acting at a point $\bm{y}$ on the surface.
Notations $P, W, O$ describe surface integrals over the ellipsoidal particle ($P$), walls ($W$) and outlet plane ($O$) respectively, as shown in Fig.~\ref{fig:schematic}(a).
$\Delta P$ is an additional pressure drop due to the presence of the particle given by
\begin{equation}
    \Delta P = \frac{1}{Q} \int_P \bm{v}^\infty \cdot \bm{q}(\bm{x}) dA
\end{equation}
where $Q$ is a flow rate.

Integrating the traction force $\bm{q}$ over the surface of the ellipsoid $P$ gives the hydrodynamic force $\bm{F}_h$ and torque $\bm{T}_h$ acting on the particle.
Since the system is force- and torque-free, these satisfy
\begin{eqnarray}
    \bm{F}_h &=& \int_P \bm{q} dA = \bm{0}, \label{eq:force} \\
    \bm{T}_h + \bm{T}_m &=& \int_P \{ \bm{q} \times (\bm{x} - \bm{x}_0) \} dA + \bm{T}_m = \bm{0} \label{eq:torque}
\end{eqnarray}
where $\bm{x}_0$ is the hydrodynamic centre of the particle.
The motion of the ellipsoid is described by 6 degrees of freedom: i.e. three translational velocities $\bm{U} = (U_x, U_y, U_z)$ and three rotational velocities $\bm{\varOmega} = (\omega_x, \omega_y, \omega_z)$.
Therefore, as the boundary condition, a given surface material point $\bm{x}_s$ on the ellipsoid moves with a velocity
\begin{equation}
    \bm{v} (\bm{x}_s) = \bm{U} + \bm{\varOmega} \times (\bm{x}_s - \bm{x}_0). \label{eq:boundary}
\end{equation}

\subsubsection{Boundary element method (BEM) \label{section:bem}}
The surface of the ellipsoid is divided into $N_E = 512$ triangular elements and $N_N = 258$ nodes, while the rectangular channel with a length $100a$ is discretised into $N_{EW} = 3920-9800$ triangular elements (wall nodes $N_{NW} = 2000-5000$) with mesh size $2a$.
In order to avoid error arising from the coarse mesh of the walls, we kept the distance between particle and walls larger than $2a$.
According to equations \eqref{eq:bie} and \eqref{eq:boundary}, the $i$-th node $\bm{x}_i$ on the particle surface has to satisfy a boundary condition
\begin{equation}
    \bm{U} + \bm{\varOmega} \times (\bm{x}_i - \bm{x}_0)
    + \frac{1}{8 \pi \eta} \left\{ \sum_{e}^{N_E + N_{EW}} \bm{G}(\bm{x}_i, \bm{y}_{e}) \bm{q} (\bm{y}_{e}) \Delta A_{e} + \bm{G}(\bm{x}_i, \bm{y}_{out}) \bm{n} \Delta P A_c \right\} = \bm{v}^\infty (\bm{x}_i) \label{eq:bem1}
\end{equation}
where $\Delta A$ is the surface area of the element, subscript $e$ is the index of elements, $\bm{y}_{e}$ is the position of the element $e$, $A_c$ is the cross-sectional area of the channel and $\bm{y}_{out}$ is a centre point in the outlet.
Similarly, the $i$-th node on the channel wall has to satisfy 
\begin{equation}
    \frac{1}{8 \pi \eta} \left\{ \sum_{e}^{N_E + N_{EW}} \bm{G}(\bm{x}_i, \bm{y}_{e}) \bm{q} (\bm{y}_{e}) \Delta A_{el} + \bm{G}(\bm{x}_i, \bm{y}_{out}) \Delta P A_c \right\} = \bm{0} \label{eq:bem2}
\end{equation}
to enforce the no-slip condition at the wall.
The force- and torque-free conditions \eqref{eq:force}-\eqref{eq:torque} can be discretised as 
\begin{eqnarray}
    \sum_{e}^{N_E} \bm{q} (\bm{x}_{e}) \Delta A_{e} &=& \bm{0}, \label{eq:force2} \\
    \sum_{e}^{N_E} \left\{ \bm{q} (\bm{x}_{e}) \times (\bm{x}_{e} - \bm{x}_0) \right\} \Delta A_{e} &=& - \bm{T}_m. \label{eq:torque2}
\end{eqnarray}
Note that 4 point Gaussian quadrature is used to calculate the surface integral over each element.
For singular elements, we work in polar coordinates to remove the $1/r$ singularity \citep{Pozrikidis1995}.
The velocities are obtained by solving the linear equations  $\bm{A} \bm{x} = \bm{b}$ with a known vector $\bm{b} = \{\bm{v}^\infty, \bm{0}, -\bm{T}_m\}$ and an unknown vector $\bm{x} = \{\bm{q}, \bm{U}, \bm{\varOmega}\}$, where $\bm{A}$ is the dense matrix of size $(3(N_N + N_{NW}) + 6)$ based on equations \eqref{eq:bem1}, \eqref{eq:bem2}, \eqref{eq:force2} and \eqref{eq:torque2}. For details, see \citet{Ishikawa2006, Hu2012, Pozrikidis2005}.  

Our goal is to use the simulations to obtain the drift velocity ($U_y, U_z$) at each position $(y, z)$ inside the channel.
Since the time scale of the rotational motion is much faster than the translational motion \citep{Matsunaga2017}, we first update the particle orientation from the angular velocity $\bm{\varOmega}$ using a 1st-order Euler method.
This process is repeated until the particle reaches the stable angles $(\theta_p^*, \phi_p^*)$, and then the drift velocities ($U_y, U_z$) are calculated.
Note that the additional pressure drop $\Delta P$ is updated every iteration by using the value $\bm{q}$ from the previous iteration. 

\begin{figure}
    \begin{center}
    \includegraphics[width=0.9\columnwidth]{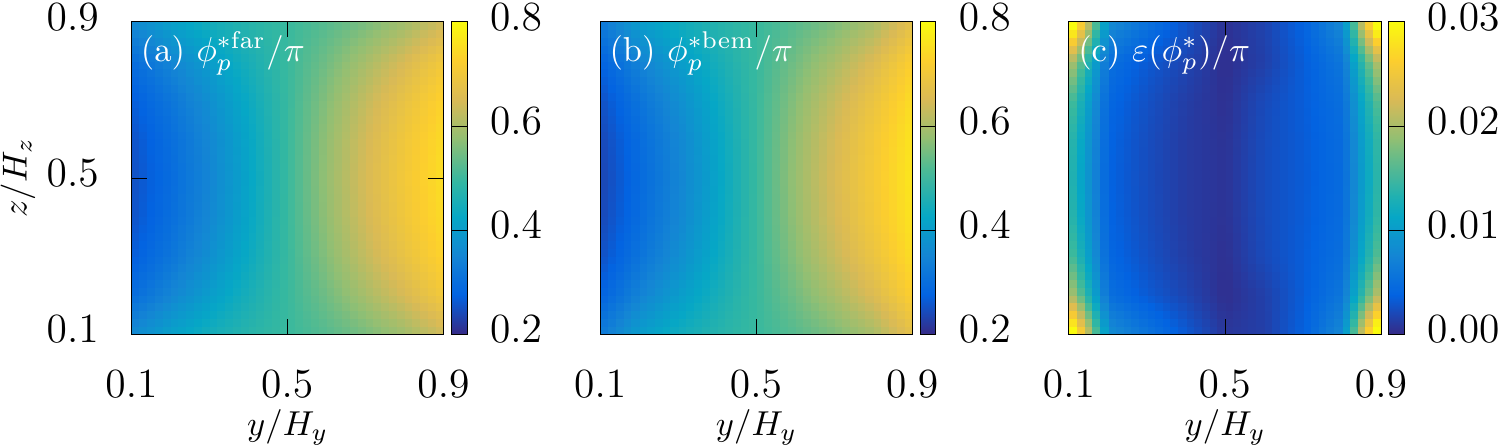}
    \includegraphics[width=0.9\columnwidth]{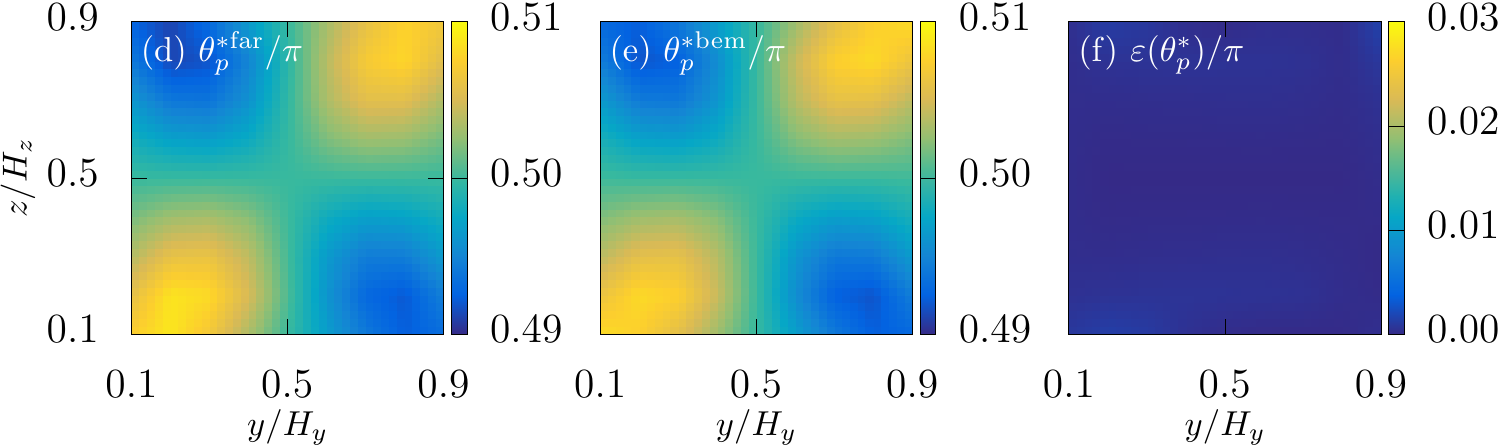}
    \end{center}
    \caption{Comparison of stable angles ($\phi_p^*, \theta_p^*$) obtained from (a),(d) the far-field theory and (b),(e) the boundary element simulations. Figures (c),(f) show the differences between the values obtained. \label{fig:compare_angle}}
\end{figure}

\subsection{Results of comparison to far-field theory}
We now compare the far-field solutions with the full simulations. We use
a particle aspect ratio $\alpha = 3$, a magnetic field $\beta_w = 30$, a magnetic field direction $\phi_B = \pi/2$ ($+y$-direction) and a channel size $H_y/a = H_z/a = 20$.

Firstly, the stable orientation angles ($\phi_p^*, \theta_p^*$) for each position in the channel are shown in Fig.~\ref{fig:compare_angle}.
Due to the strong pinning by the external field, $\beta_w = 30$, the particle orientation is almost in the ($x, y$)-plane, $\theta_p^* \approx \pi/2$, as shown in Figs.~\ref{fig:compare_angle}(d),(e). 
At the four corners of the channel, the particles are slightly tilted out of the ($x, y$)-plane because of local shear forces.
On the other hand, $\phi_p^*$ monotonically increases with $y$ as shown in Figs.~\ref{fig:compare_angle}(a),(b) because of the balance between the local shear $\dot{\gamma}_{xy}$ and the magnetic torque.
For both angles, the far-field theory reproduces the distribution obtained from the full simulation well.
To measure the differences 
\begin{equation}
    \varepsilon (X) = | X^{\rm far} - X^{\rm bem} |,
\end{equation}
where $X^{\rm far}$ and $X^{\rm bem}$ are the values of a given variable $X$ obtained from the far-field theory and the BEM simulations respectively, are also plotted. The errors are only up to a few percent.
The good agreement is because the contribution of the image rotlet quickly decays, with $O(h^3)$, where $h$ is distance between the wall and the particle \citep{Spagnolie2012}.
Small deviations of $\varepsilon(\phi_p^*)$ at the channel corners are due to omitting the higher-order terms of the flow field, such as shear gradients across the particle, in the far-field theory.

\begin{figure}
    \begin{center}
    \includegraphics[width=0.9\columnwidth]{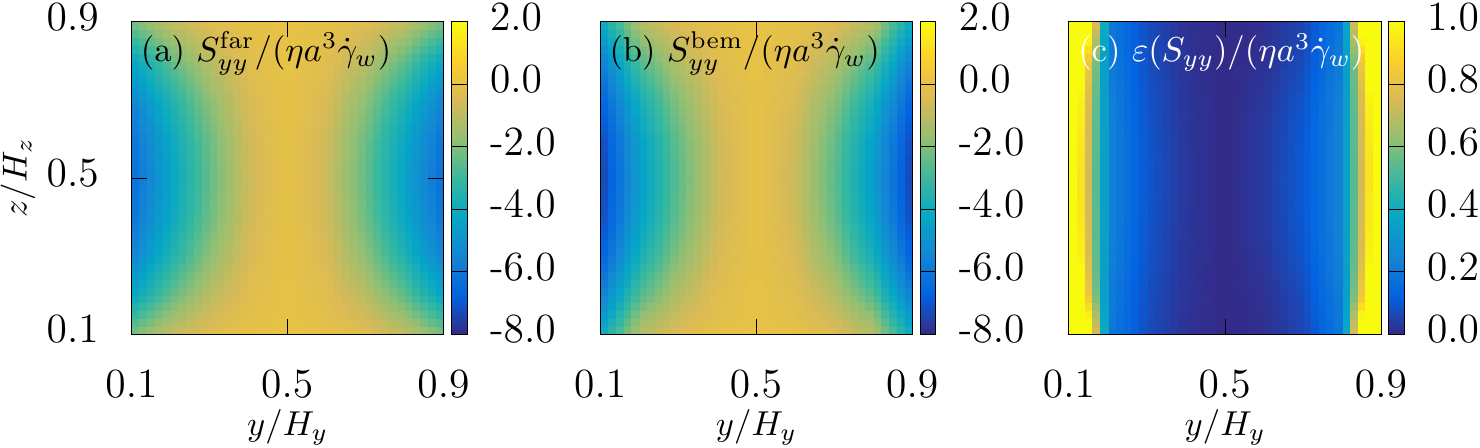}
    \includegraphics[width=0.9\columnwidth]{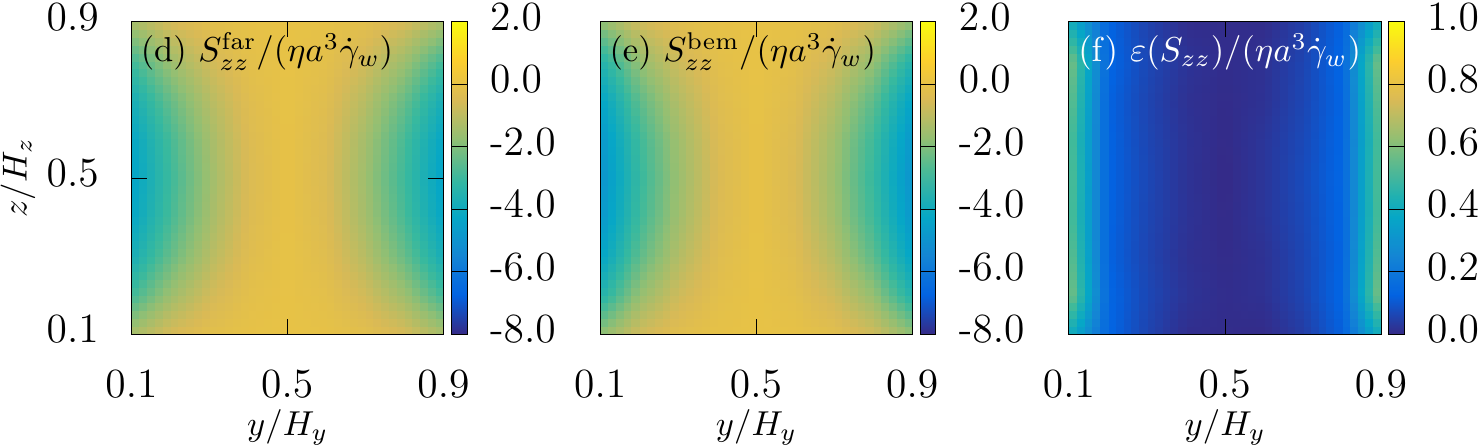}
    \end{center}
    \caption{Comparison of the stresslets ($S_{yy}$, $S_{zz}$) obtained from (a),(d) the far-field theory and (b),(e) the boundary element simulations. Figures (c),(f) show the differences between the values obtained. \label{fig:compare_stresslet}}
\end{figure}

Secondly, in Fig.~\ref{fig:compare_stresslet}, we compare the stresslet ($S_{yy}$, $S_{zz}$) distributions inside the channel.
For the analytical calculations,  we use $S_{yy}^\infty$ and $S_{zz}^{\infty}$ defined in Eqs. \eqref{eq:Syy} and \eqref{eq:Szz}.
For the boundary element simulations the stresslets were calculated using
\begin{equation}
    S_{ij} = \sum_{e}^{N_E} \left\{ x_i q_j  + x_j q_i - \frac{2}{3} \delta_{ij} x_k q_k \right\} \Delta A_{e}
\end{equation}
based on the definition in \citet{KimBook}.
We also obtained results for $S_{yz}$ showing that the values are one order of magnitude smaller than $S_{yy}$ and $S_{zz}$ (data not shown).
The far-field theory reproduces the simulation results well except close to the walls:
at the channel corners, the error of the far-field theory reaches up to 40\%. This is because 
we ignored the effect of the walls modifying the stresslet. 

\begin{figure}
    \begin{center}
    \includegraphics[width=0.6\columnwidth]{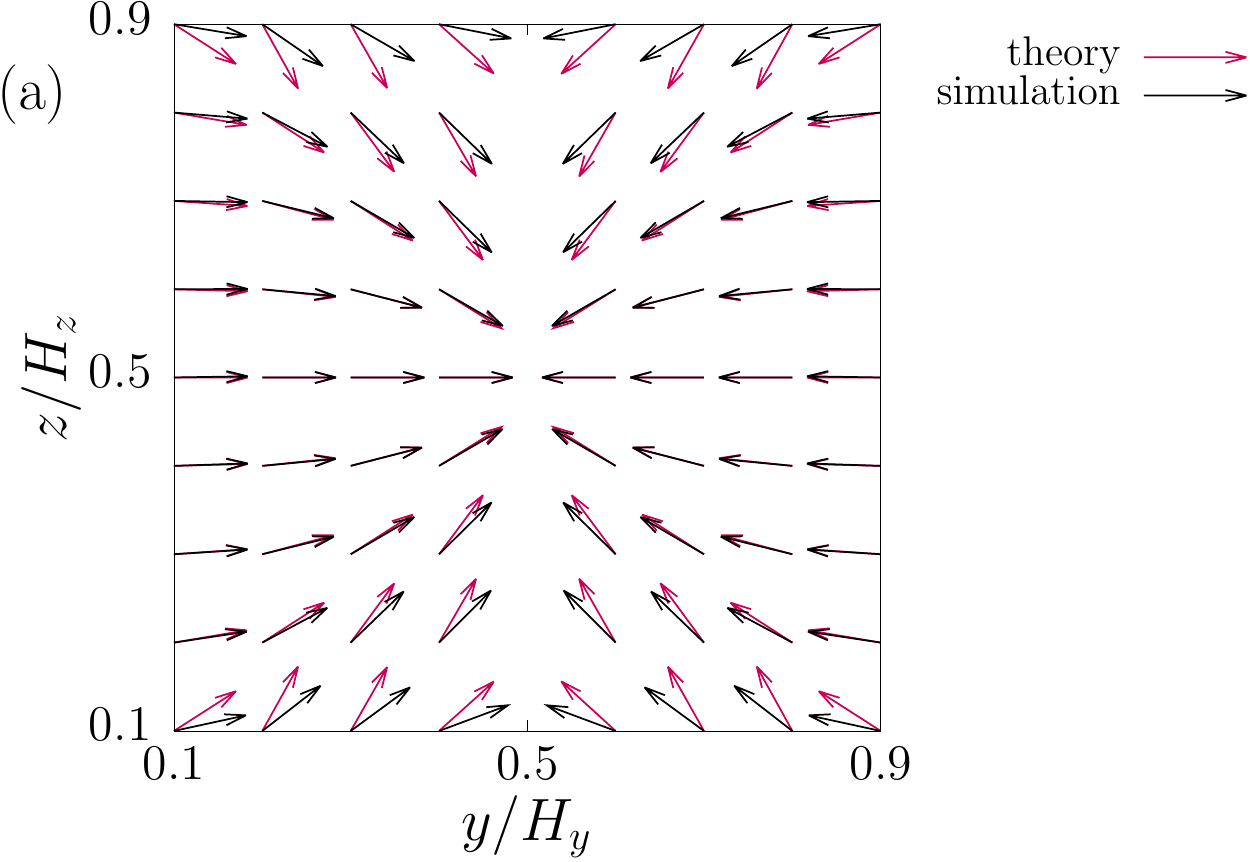}
    \includegraphics[width=\columnwidth]{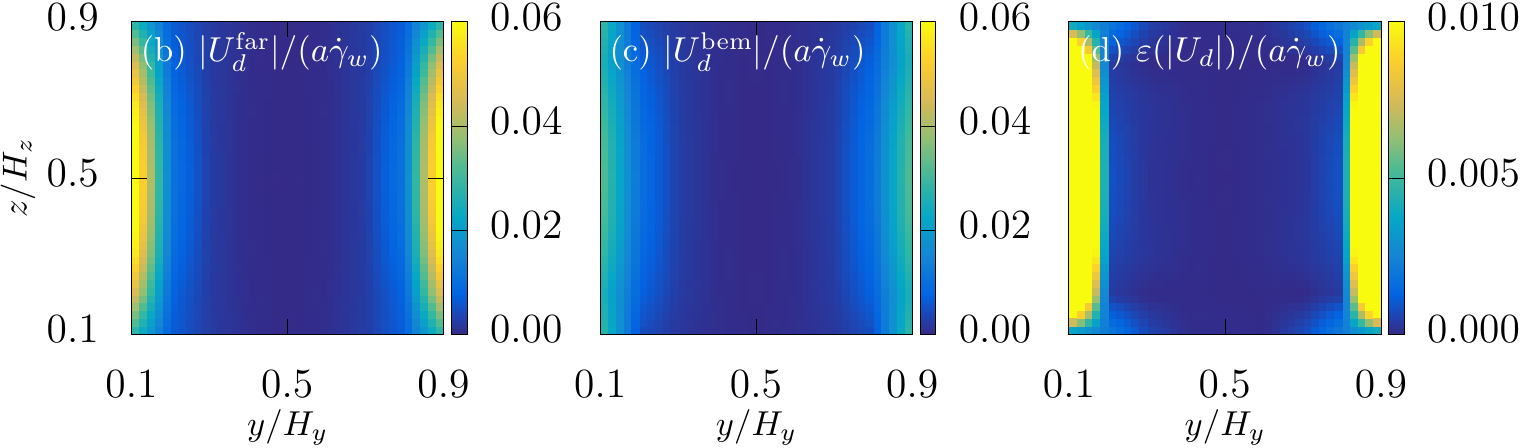}
    \end{center}
    \caption{Comparison of the particle drift velocity $\bm{U}_d$ obtained from the far-field theory and the boundary element simulations. (a) Direction of  $\bm{U}_d$. (b)--(d) Velocity amplitude $| \bm{U}_{d} | = \sqrt{U_y^2 + U_z^2}$ from (b) the far-field theory, (c) the boundary element simulations. (d) Difference between the two values. \label{fig:compare_velocity}}
\end{figure}

Next, we compare the particle drift velocity inside the channel.
Figure \ref{fig:compare_velocity}(a) shows the direction of the particle drift velocity ($U_y, U_z$) over a channel cross-section.
Although there are slight differences in the drifting directions, the quantitative character of the movements are well captured by the far-field theory.
On the other hand, the amplitude of the drift velocity $|\bm{U}_{d}| = \sqrt{U_y^2 + U_z^2}$ has large deviations $\varepsilon(| \bm{U}_{d} |)$, of up to 90\% when the particles are close to the walls at $y/H_y=0$ and $1$,
as shown in Figs.~\ref{fig:compare_velocity}(b)-(d).
This large quantitative error is due to using the free space stresslet, and also to the absence of the quadrupole and higher order image multipoles in the theory.

\begin{figure}
    \begin{minipage}{0.47 \columnwidth} \centerline{\begin{overpic}[width=\columnwidth]{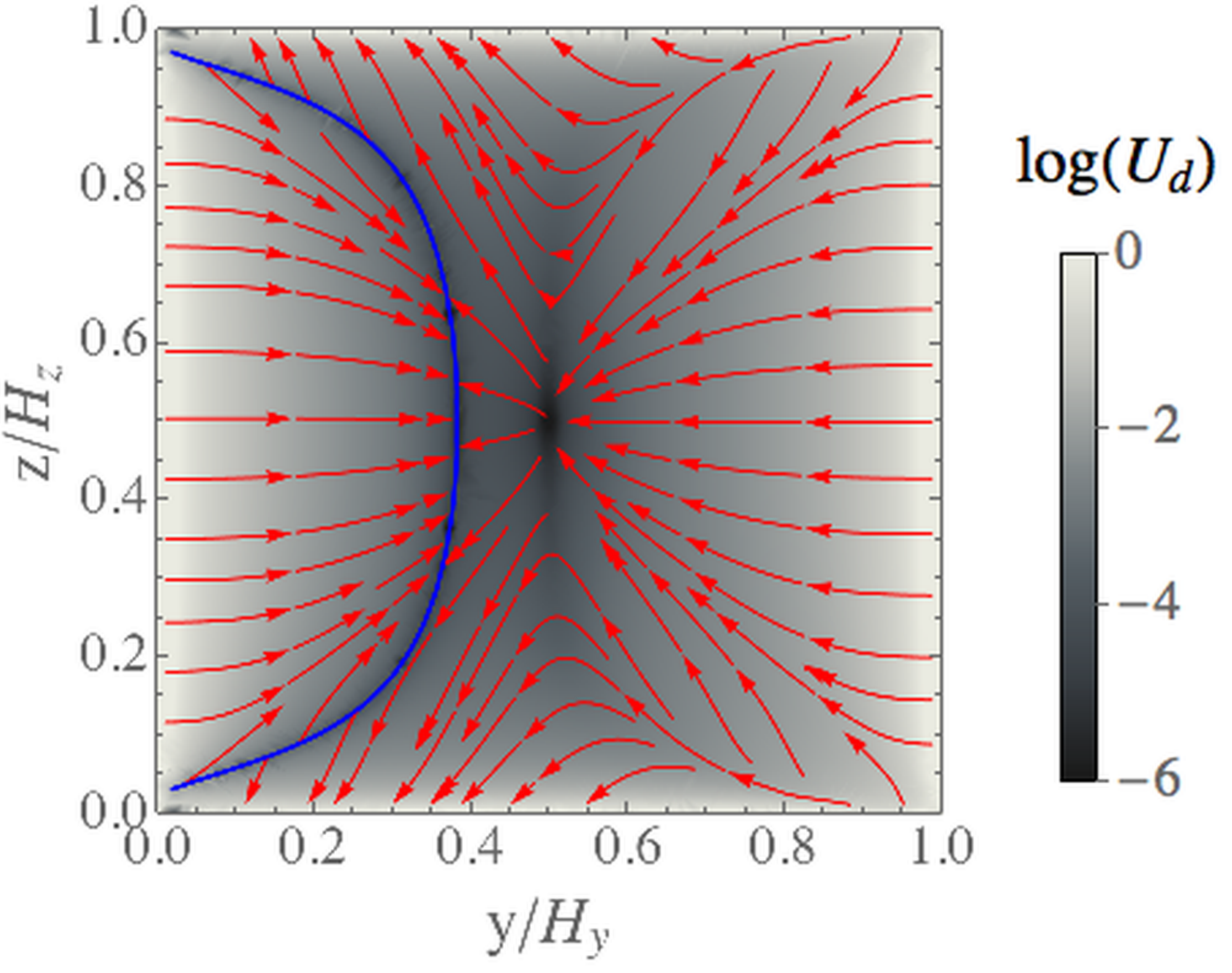} \put(-1,70){(a)}\end{overpic}} \end{minipage}
    \begin{minipage}{0.47 \columnwidth} \centerline{\begin{overpic}[width=\columnwidth]{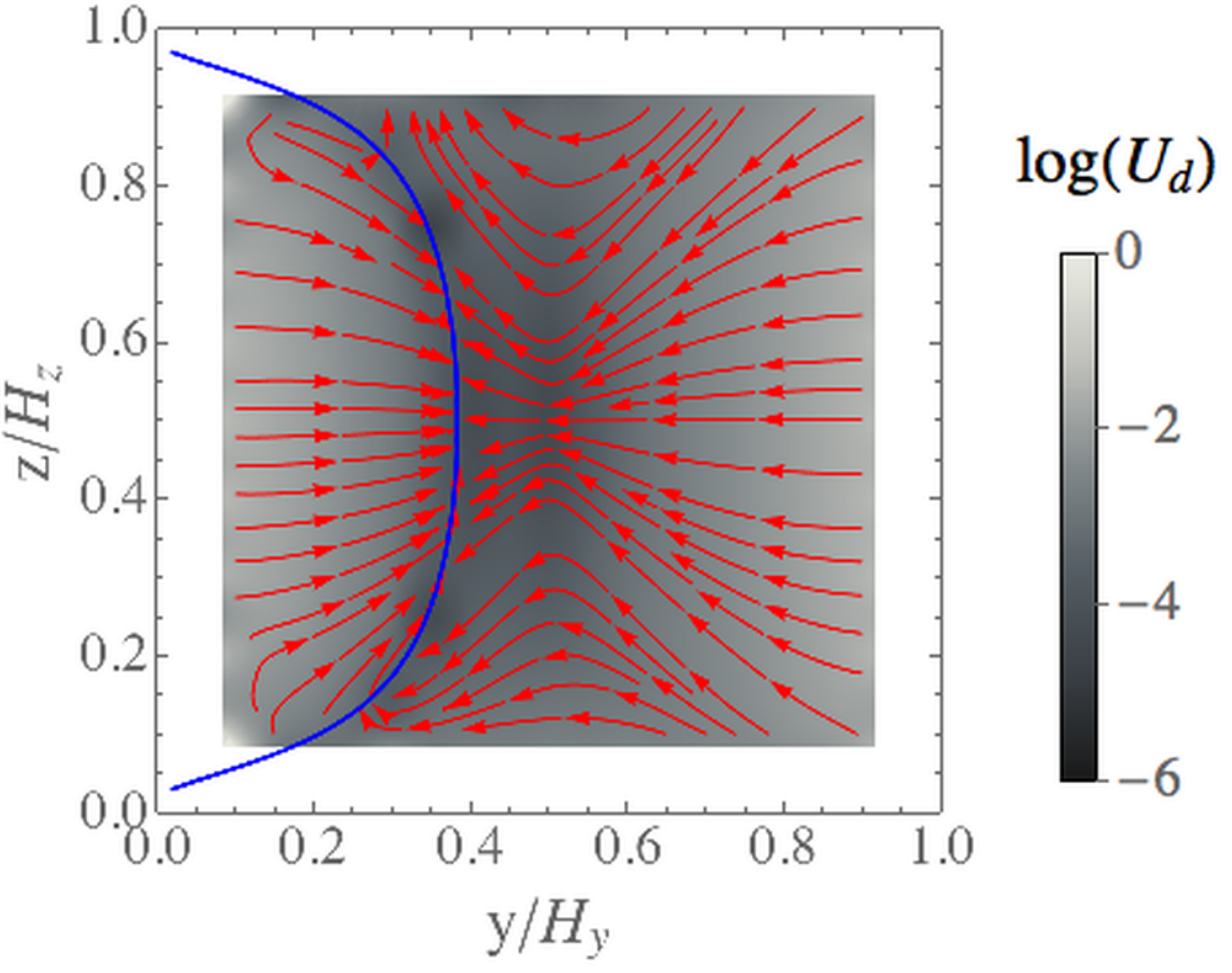} \put(-1,70){(b)}\end{overpic}} \end{minipage}
        \caption{Comparison of the particle trajectories obtained from (a) the far-field theory and (b) the boundary element simulations. Parameters are $\alpha = 3$, $\beta_w = 30$, $\phi_B = -0.4\pi$ and $H_y/a = H_z/a = 20$. Red lines follow the particle paths, while grey contours show the in-plane velocity amplitudes $\bm{U}_{d}$. Blue lines in (a) and (b) indicate the focusing region $|\bm{U}_d| = 0$, obtained from the theory. \label{fig:streamplot0}}
\end{figure}

Finally in Fig.~\ref{fig:streamplot0}, we compare the migration and destinations from the theory and the simulations showing excellent agreement.
Blue lines in the figure indicate the focusing region $U_d = 0$, and there is a perfect match.
Thus deviations in the velocity magnitude near the walls are unimportant in determining the final particle destinations, and we will discuss the particle trajectories based on the far-field theory in the next section.

\section{Results: focusing in rectangular channels}
Although the boundary element simulation provides the velocity field without approximations, this approach is not suitable for systematic parameter scanning because of the heavy computational costs.
On the other hand the far-field theory, validated in the previous section, is a quick and robust method to estimate the particle motion.
Therefore we now report details of the particle motion inside the channel based on the far-field theory.
We concentrate in particular on the particle trajectories, destinations and focusing regions \citep{Matsunaga2017}.

\begin{figure}
    \begin{minipage}{0.47 \columnwidth} \centerline{\begin{overpic}[width=\columnwidth]{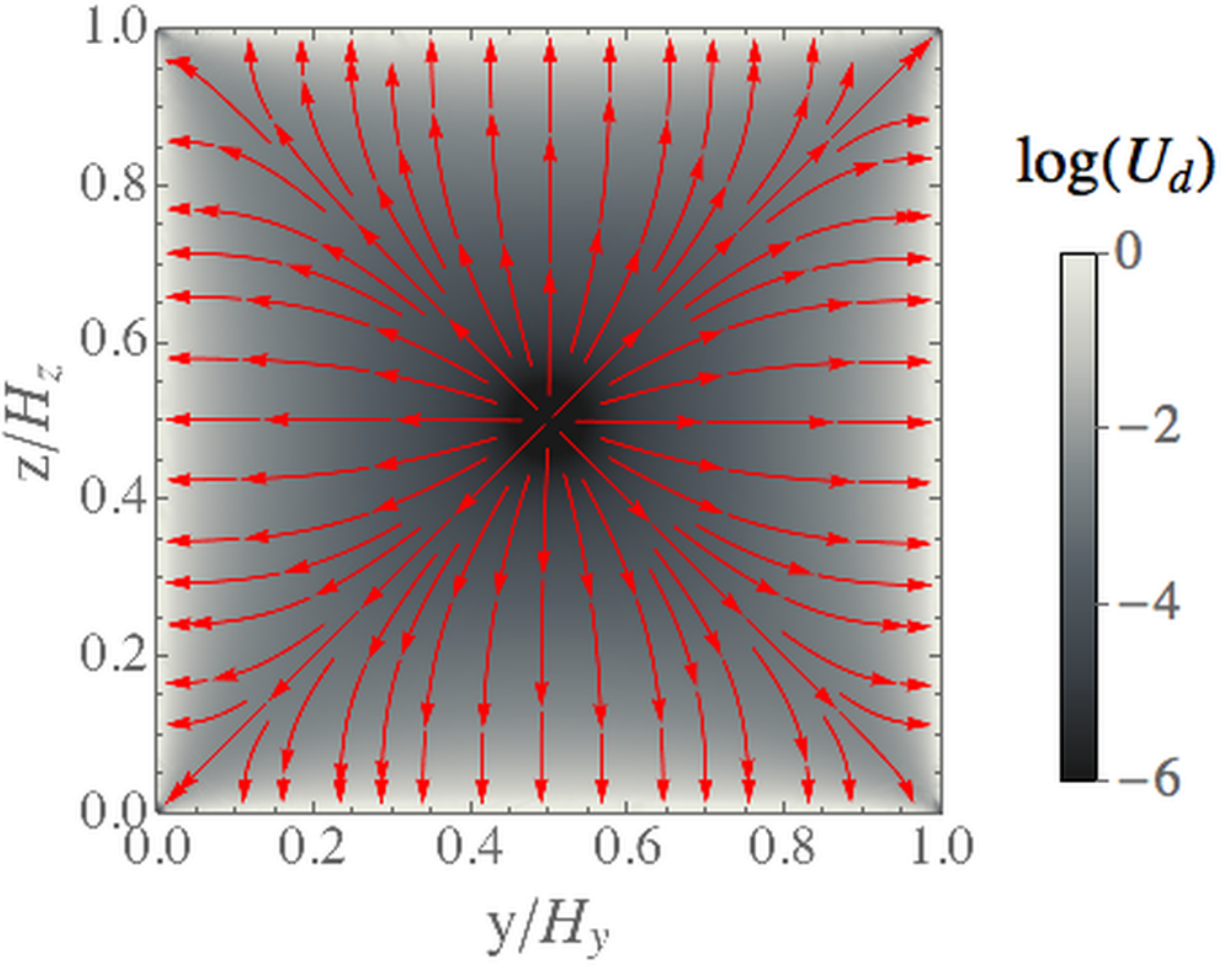} \put(-1,70){(a)}\end{overpic}} \end{minipage}
    \begin{minipage}{0.47 \columnwidth} \centerline{\begin{overpic}[width=\columnwidth]{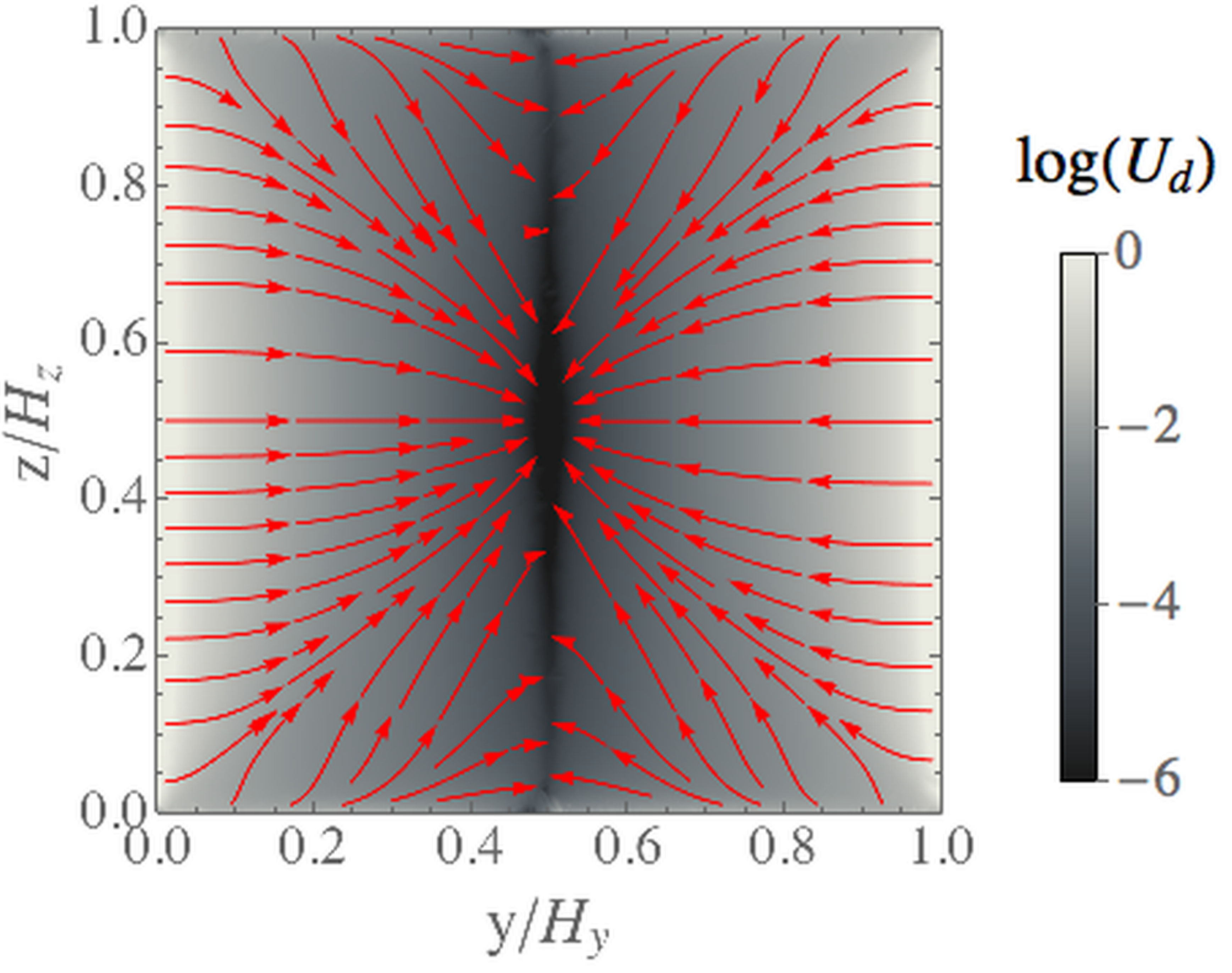} \put(-1,70){(b)}\end{overpic}} \end{minipage}
    \caption{Particle trajectories projected onto the $y-z$ plane for $\alpha = 3$, $H_y/a = H_z/a = 20$, $\beta_w = 60$, and (a) $\phi_B = 0$ and (b) $\pi/2$. \label{fig:streamplot1}}
\end{figure}

\subsection{Particle destination}
First, in Fig.~\ref{fig:streamplot1}, we show particle trajectories projected onto the $y-z$ plane for the two simplest cases: the magnetic field $\bm{B}$ pointing along the flow direction ($\phi_B = 0$) and perpendicular to the flow direction ($\phi_B = \pi/2$), for parameter values $\alpha = 3$, $\beta_w = 60$ and $H_y/a = H_z/a = 20$.
For $\phi_B = 0$ the particles all move towards the closest wall. For $\phi_B = \pi/2$, however, they flow towards $y/H_y = 0.5$ and, after
a sufficiently long time, focus to the $y/H_y = 0.5$ plane, as indicated by the black contour in Fig.~\ref{fig:streamplot1}(b).
A similar behaviour was observed for $\phi_B = 0$, $\pi/2$ for confinement between two infinite walls \citep{Matsunaga2017}.
In both cases, the particles move more quickly when they are closer to the walls because the drift velocity is provided by the image stresslets \citep{Blake1971}.

\begin{figure}
    \begin{center}
    \begin{minipage}{0.490 \columnwidth} \centerline{\begin{overpic}[width=\columnwidth]{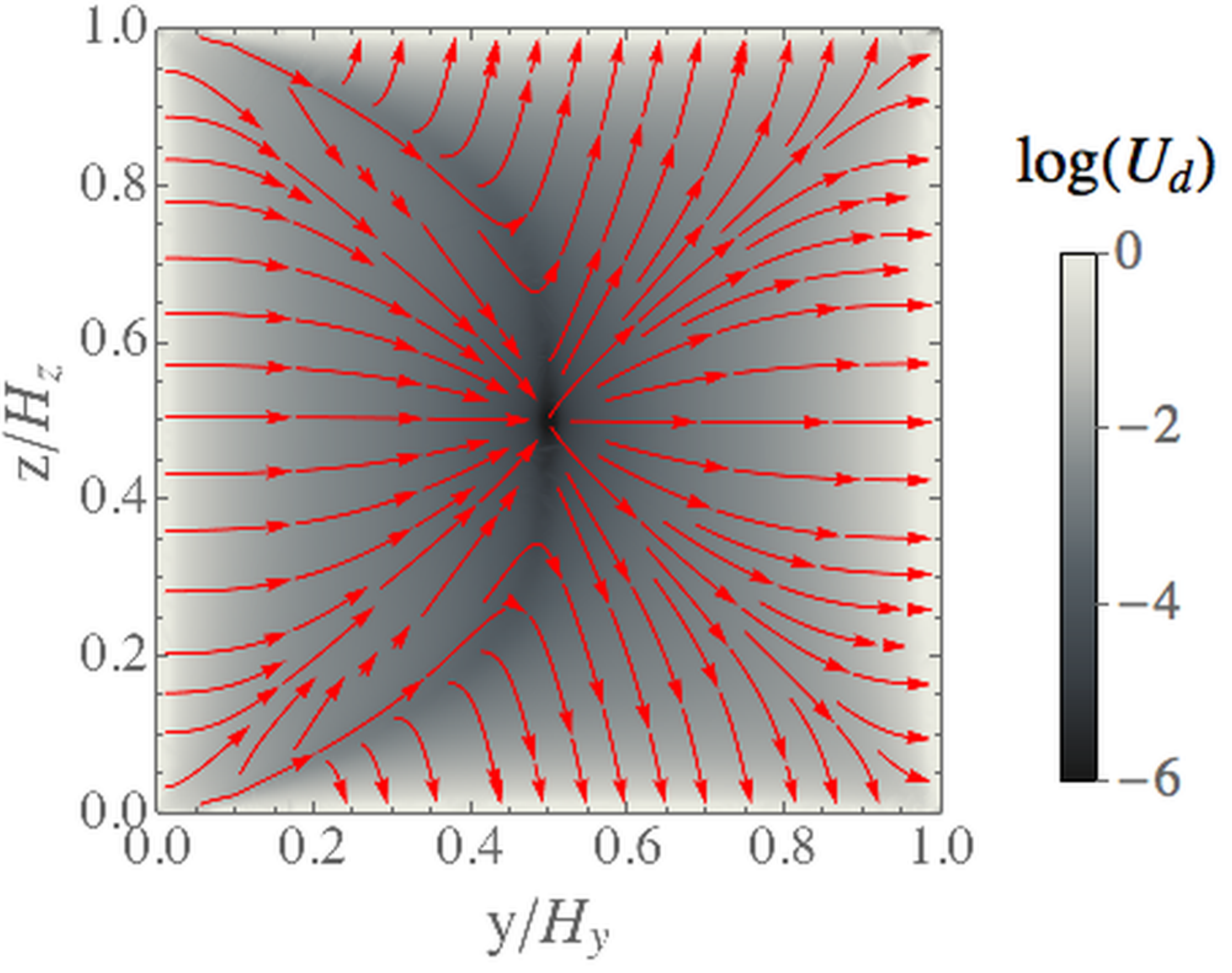} \put(-1,70){(a)}\end{overpic}} \end{minipage}
    \begin{minipage}{0.490 \columnwidth} \centerline{\begin{overpic}[width=\columnwidth]{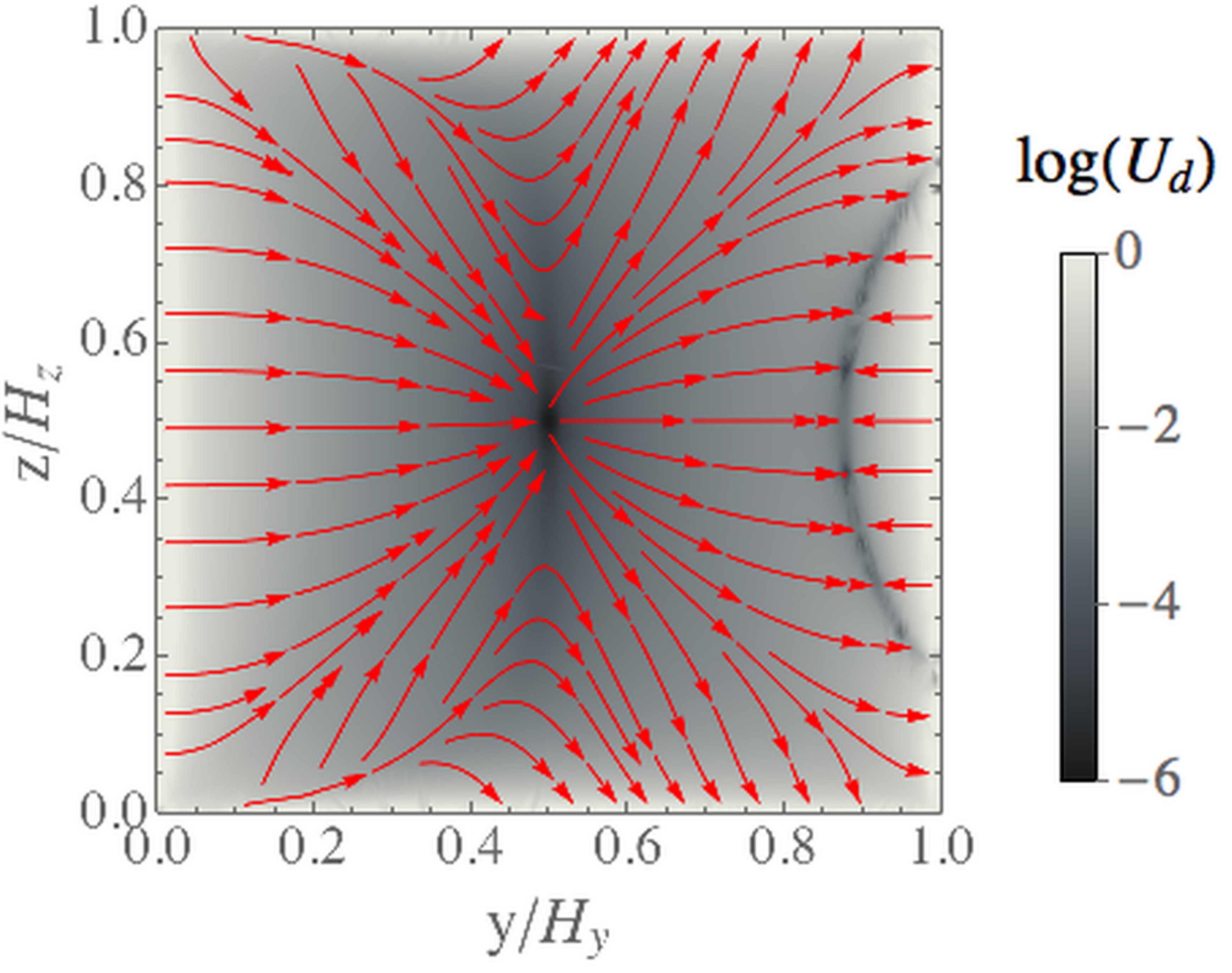} \put(-1,70){(b)}\end{overpic}} \end{minipage}
    \begin{minipage}{0.490 \columnwidth} \centerline{\begin{overpic}[width=\columnwidth]{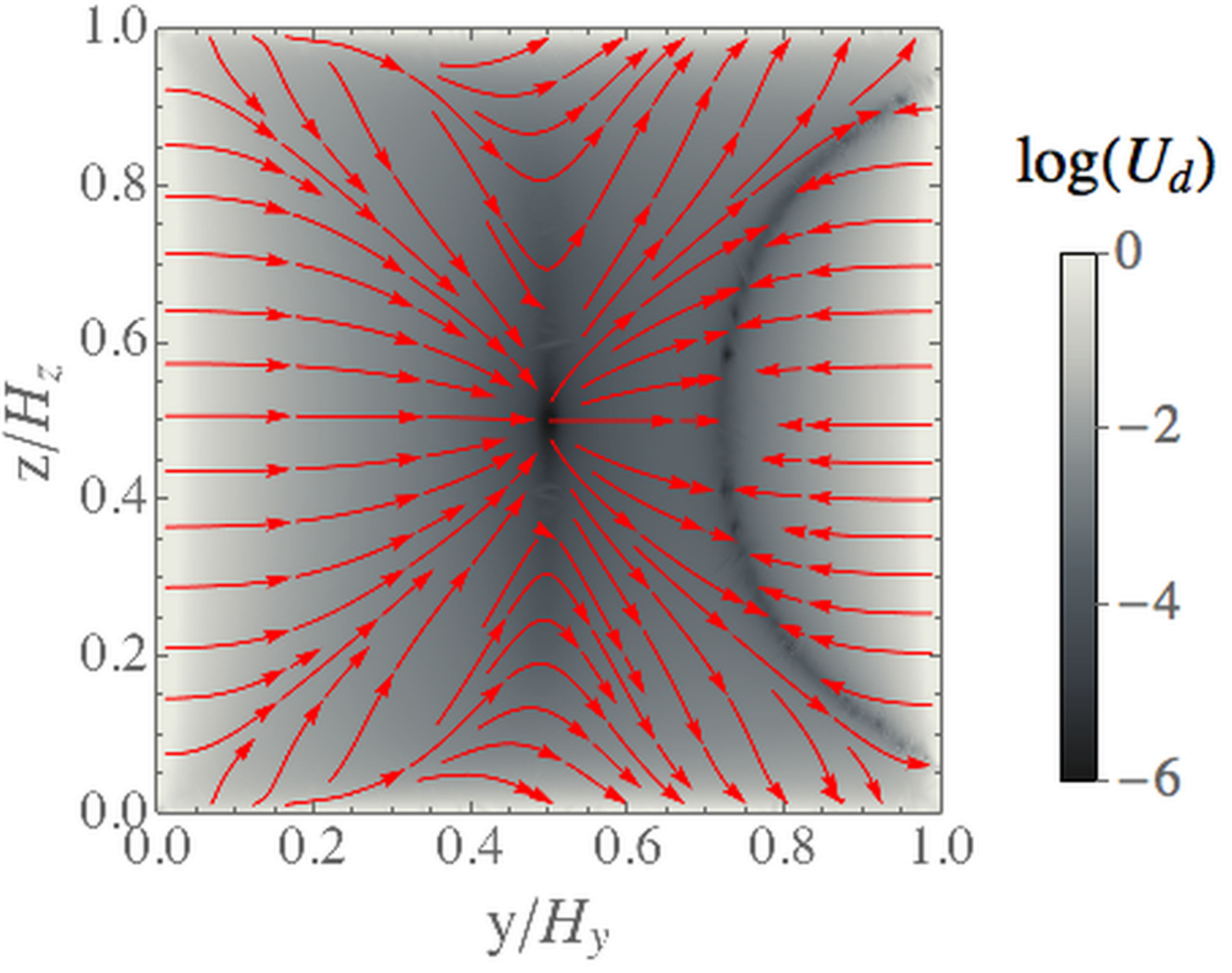} \put(-1,70){(c)}\end{overpic}} \end{minipage}
    \begin{minipage}{0.490 \columnwidth} \centerline{\begin{overpic}[width=\columnwidth]{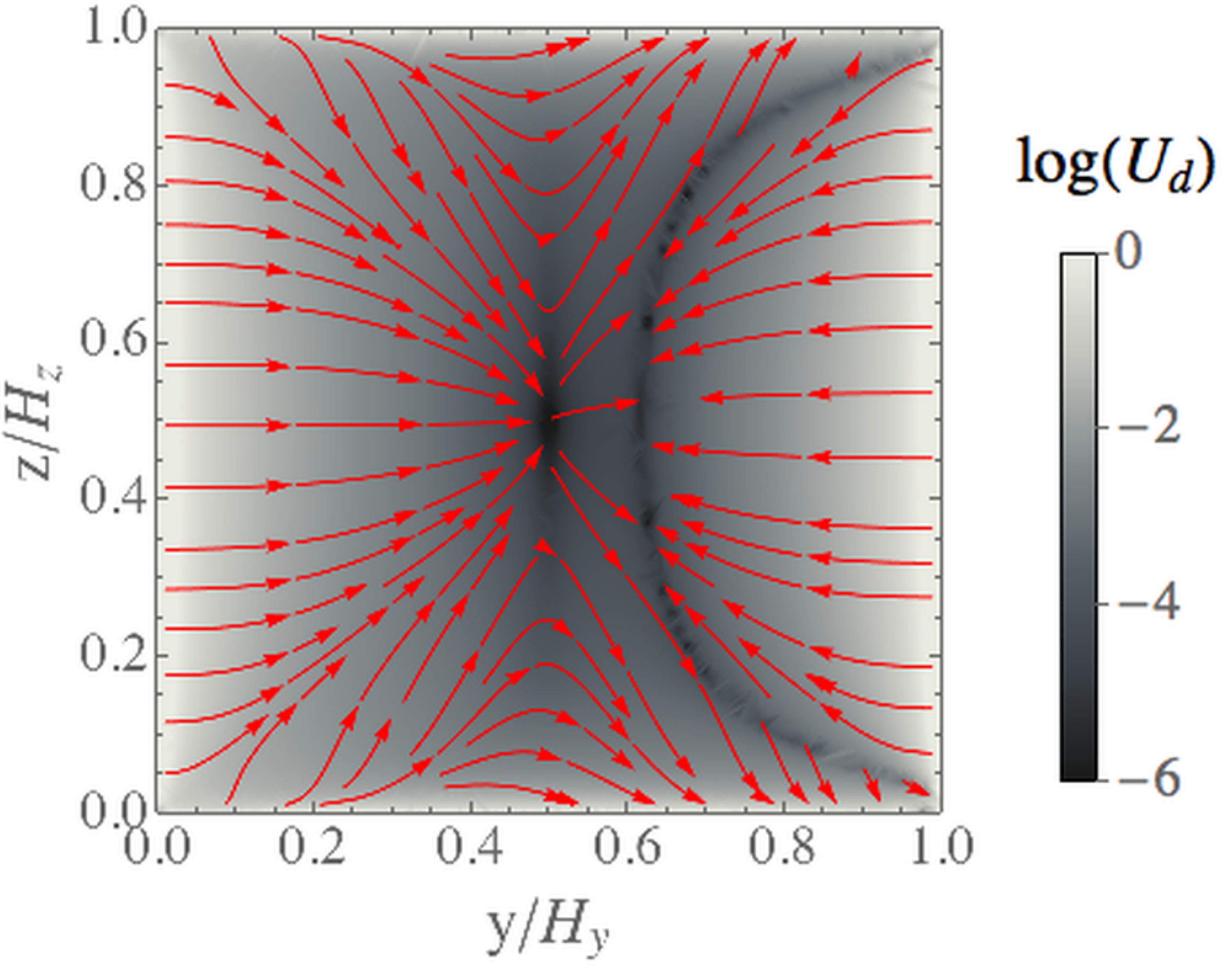} \put(-1,70){(d)}\end{overpic}} \end{minipage}
    \end{center}
    \caption{Particle trajectories projected onto the $y-z$ plane for $\alpha = 3$, $H_y/a = H_z/a = 20$, $\beta_w = 60$, and (a) $\phi_B = 0.10\pi$, (b) $0.30\pi$, (c) $0.40\pi$ and (d) $0.45\pi$. \label{fig:streamplot2}}
\end{figure}

When the direction of the magnetic field is changed to other orientations $0 < \phi_B < \pi/2$, the trajectories of the particles become more complex as shown in Fig.~\ref{fig:streamplot2}. The majority of the particles move to one of three destinations, depending on their initial positions:
\begin{itemize}
    \item{(i) curved focusing regions indicated with black contours,}
    \item{(ii) the walls at $z/H_z=0$ or $z/H_z=1$, or}
    \item{(iii) the wall at $y/H_y=1$.}
\end{itemize}
By changing the field direction $\phi_B$, we can change not only the position of the focusing region (i) but also the ratio of particles reaching each destination (i)-(iii).
When the applied magnetic field is close to the flow direction as in Fig.~\ref{fig:streamplot2}(a), most of the particles reach regions (ii) or (iii), ie they move away from the wall at $y/H_y=0$ towards the other three walls.
When the applied field is closer to $\phi_B = \pi/2$ however, as in Fig.~\ref{fig:streamplot2}(d), most of the particles reach regions (i) or (ii). 
For intermediate angles, Figs.~\ref{fig:streamplot2}(b),(c), the particles are distributed over all three destinations.

As reported in our previous study \citep{Matsunaga2017}, particles with different shapes can be sorted using region (i) because the focusing positions are dependent on the particle aspect ratios $\alpha$.
There are two possible ways to focus all the particles to the curved focusing region (i) without losing them to walls (ii)-(iii).
The first method is to control the initial position of the particles.
By designing the particle inlet position to be within a region between the curved line (i) and the wall at $y/H_y=1$, the particle will always be focused to region (i). The second method is to increase the aspect ratio of the channel, $H_z/H_y$, as we discuss in the next subsection. 

\subsection{Focusing region: Effect of parameters}
Under confinement between two infinite walls, $H_z \to \infty$ \citep{Matsunaga2017}, the particle is focused to a stable position
\begin{equation}
    \frac{y*}{H_y} (\alpha, \beta_w, \phi_B) = \frac{1}{2} \pm \frac{\beta_w F(\alpha)}{8 \pi (1 + J(\alpha))} \cos \phi_B, \label{eq:focusing}
\end{equation}
which depends on the particle aspect ratio $\alpha$, the magnetic field strength $\beta_w$ and the direction of the external field $\phi_B$.
For a rectangular channel, the focusing region  also depends on the aspect ratio of the channel $H_z/H_y$.
In this subsection, we show how each parameter affects the focusing region (i).

\begin{figure}
    \begin{center}
    \begin{minipage}{0.32 \columnwidth} \centerline{\includegraphics[width=\columnwidth]{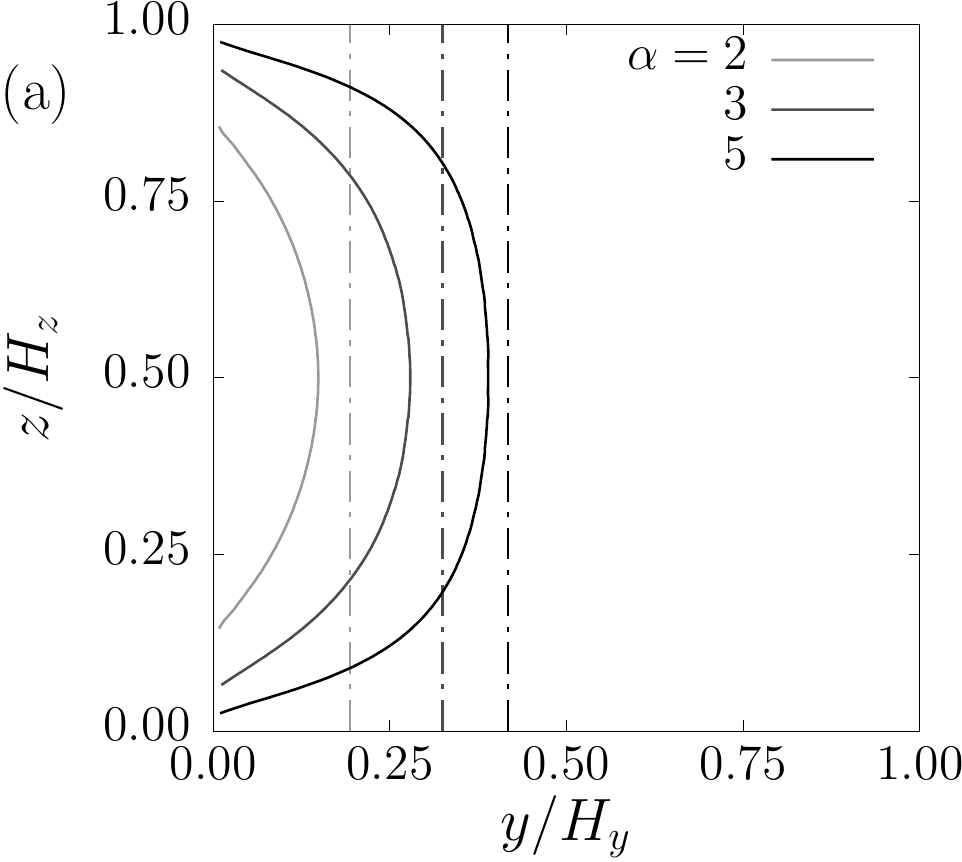}} \end{minipage}
    \begin{minipage}{0.32 \columnwidth} \centerline{\includegraphics[width=\columnwidth]{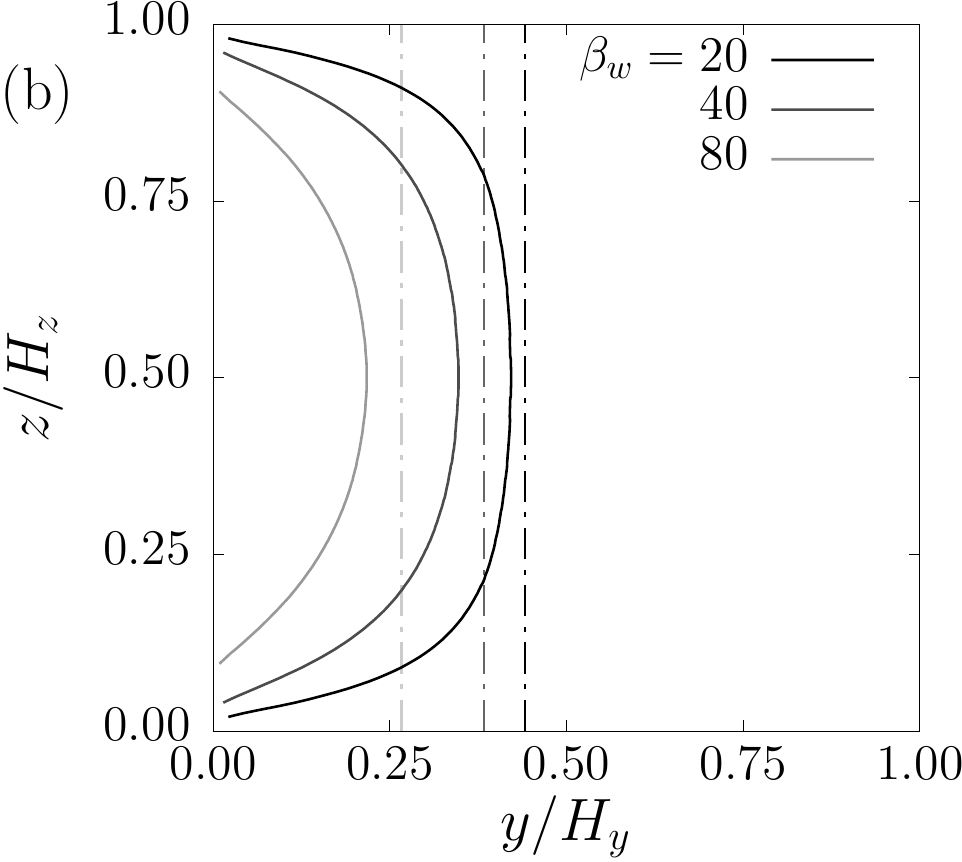}} \end{minipage}
    \begin{minipage}{0.32 \columnwidth} \centerline{\includegraphics[width=\columnwidth]{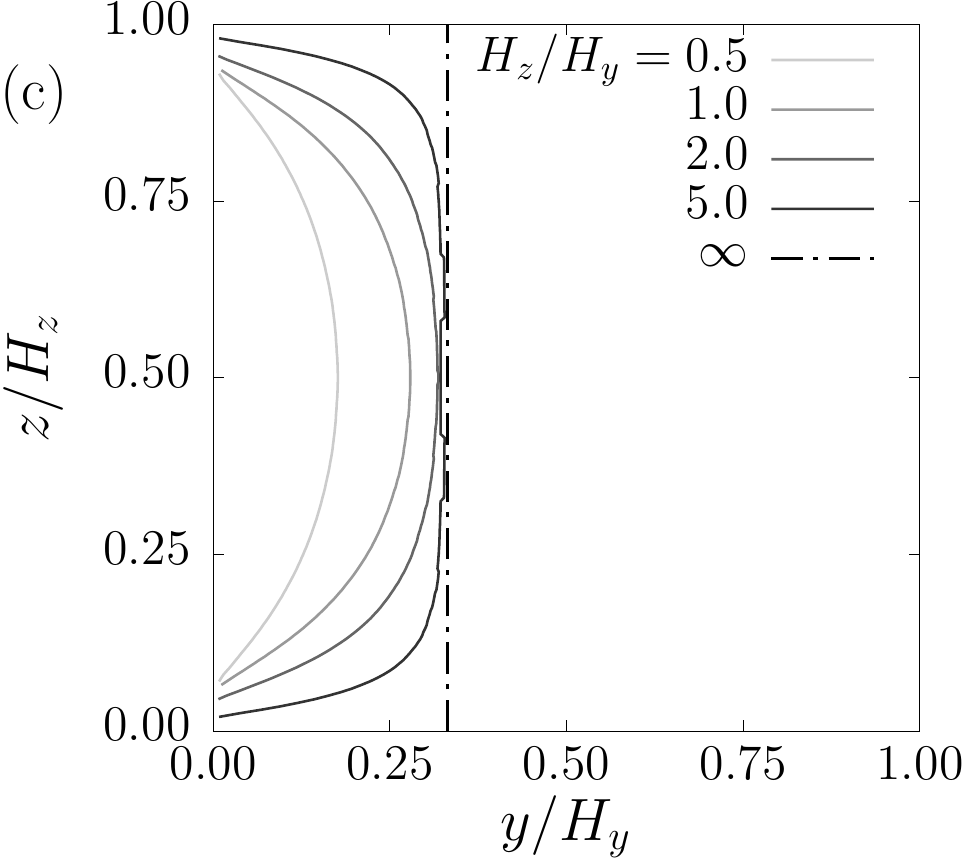}} \end{minipage}
    \end{center}
    \caption{Comparison of the focusing region under different conditions: Effect of (a) the particle aspect ratio $\alpha$, (b) the magnetic field $\beta_w$ and (c) the aspect ratio of the channel $H_z/H_y$. Conditions are $\alpha = 3$, $\beta_w = 60$, $\phi_B = -0.4$ and $H_y/a = H_z/a = 20$ if the value is not otherwise specified. Dotted vertical lines show the focusing line \eqref{eq:focusing} for confinement between two infinite walls $H_z \to \infty$. \label{fig:focusing}}
\end{figure}

Fig.~\ref{fig:focusing}(a) compares the position and shape of the focusing region for particles with different aspect ratios $\alpha$.
As predicted by Eq.~\eqref{eq:focusing}, particles with smaller aspect ratio are focused closer to the wall at $y/H_y=0$.
Hence particles with different aspect ratios can be sorted.
Fig.~\ref{fig:focusing}(b) considers the effect of the strength of the magnetic field $\beta_w$, showing that a stronger field also leads to focusing
closer to the wall.
By changing both $\beta_w$ and $\phi_B$, the focusing region can easily be controlled.
Finally, Fig.~\ref{fig:focusing}(c) compares the focusing region for channels with different aspect ratios $H_z/H_y$.
Note that the length $H_z$ is modified to change the aspect ratio, while $H_y/a = 20$ is set constant.
The particles are focused to a region of higher curvature for smaller $H_z/H_y$ while they are focused to a line for $H_z/H_y \to \infty$.
Hence in order to focus the particles to the vicinity of a chosen value of $y$, the aspect ratio of the channel should be sufficiently large.

\section{Results: Focusing in circular channels}
As is apparent from our results in the previous sections, the directions in which the particles move are strongly dependent on the channel geometry because the drift velocities are given by the image stresslets.
Therefore, it is interesting to study the focusing under different confinements, and
in this section we briefly describe focusing in circular channels.
As far as we are aware an analytic expression for the image stresslet in circular channels is not available so we only present results from numerical simulations.

We use the same numerical method as for the  channel of rectangular cross section, described in section \ref{section:bem}, 
 except for the flow field and the channel shape. 
The background velocity inside a circular channel with radius $R$ is
\begin{equation}
    v_x^\infty (r) = \frac{R \dot{\gamma}_w}{2} \{ 1 - (r/R)^2 \}
\end{equation}
where $r$ is the radial position and $\dot{\gamma}_w$ is the wall shear rate.
The channel has length $100a$ and is discretized into triangular elements with mesh size $a$.
To avoid errors arising from the coarse mesh of the walls, we kept the distance between particle and walls larger than $2a$.
\begin{figure}
    \begin{center}
    \begin{minipage}{0.490 \columnwidth} \centerline{\begin{overpic}[width=\columnwidth]{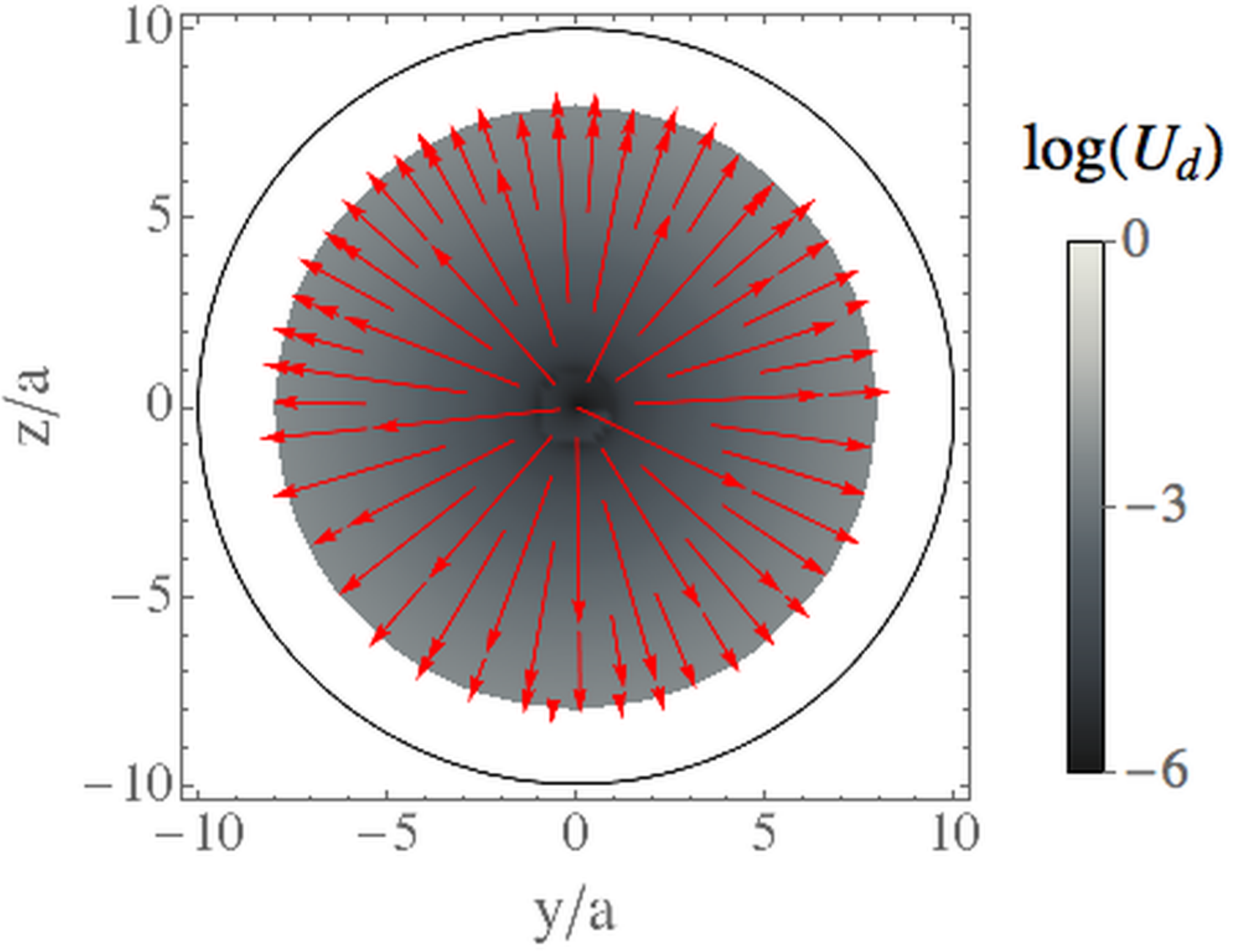} \put(-1,70){(a)}\end{overpic}} \end{minipage}
    \begin{minipage}{0.490 \columnwidth} \centerline{\begin{overpic}[width=\columnwidth]{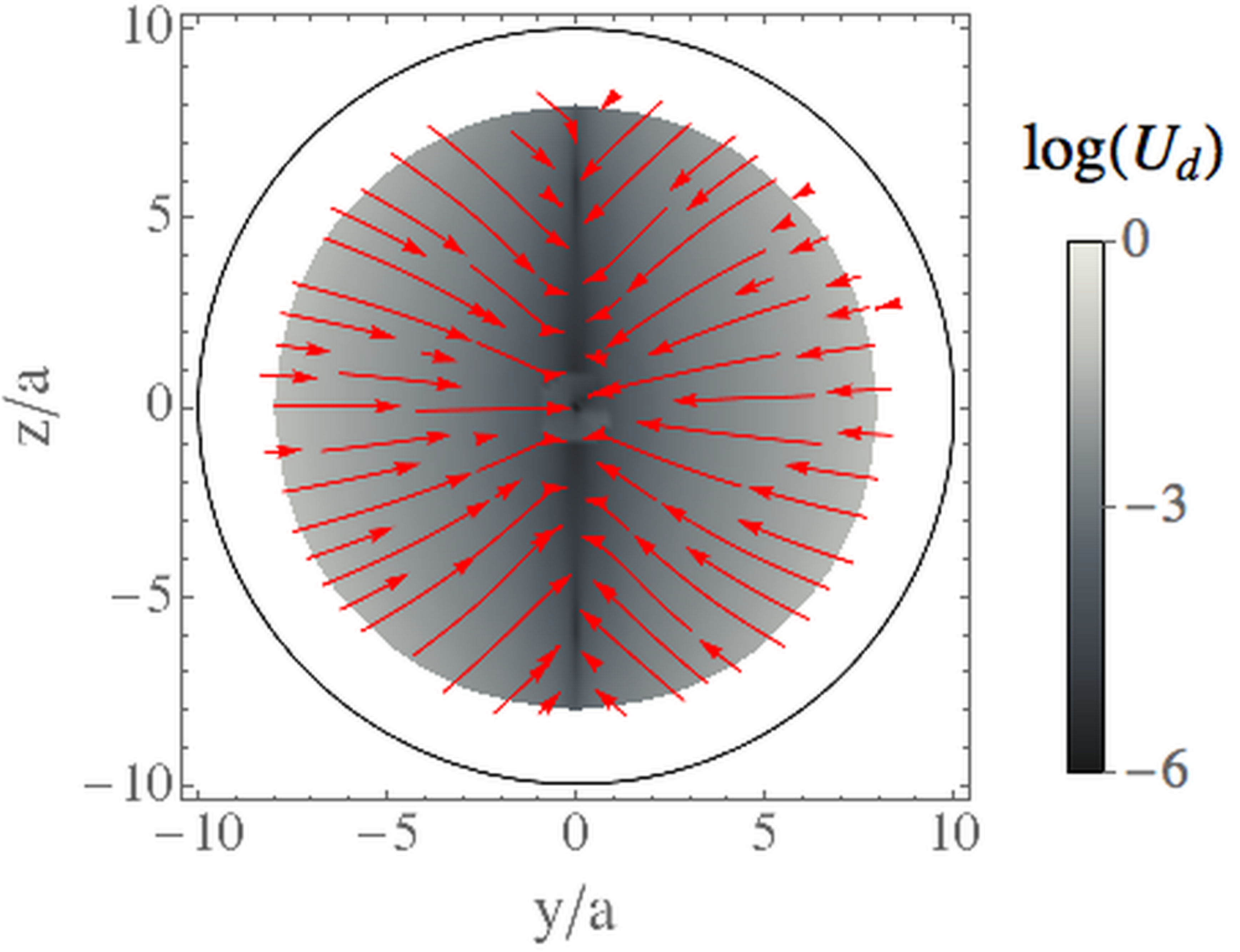} \put(-1,70){(b)}\end{overpic}} \end{minipage}
    \begin{minipage}{0.490 \columnwidth} \centerline{\begin{overpic}[width=\columnwidth]{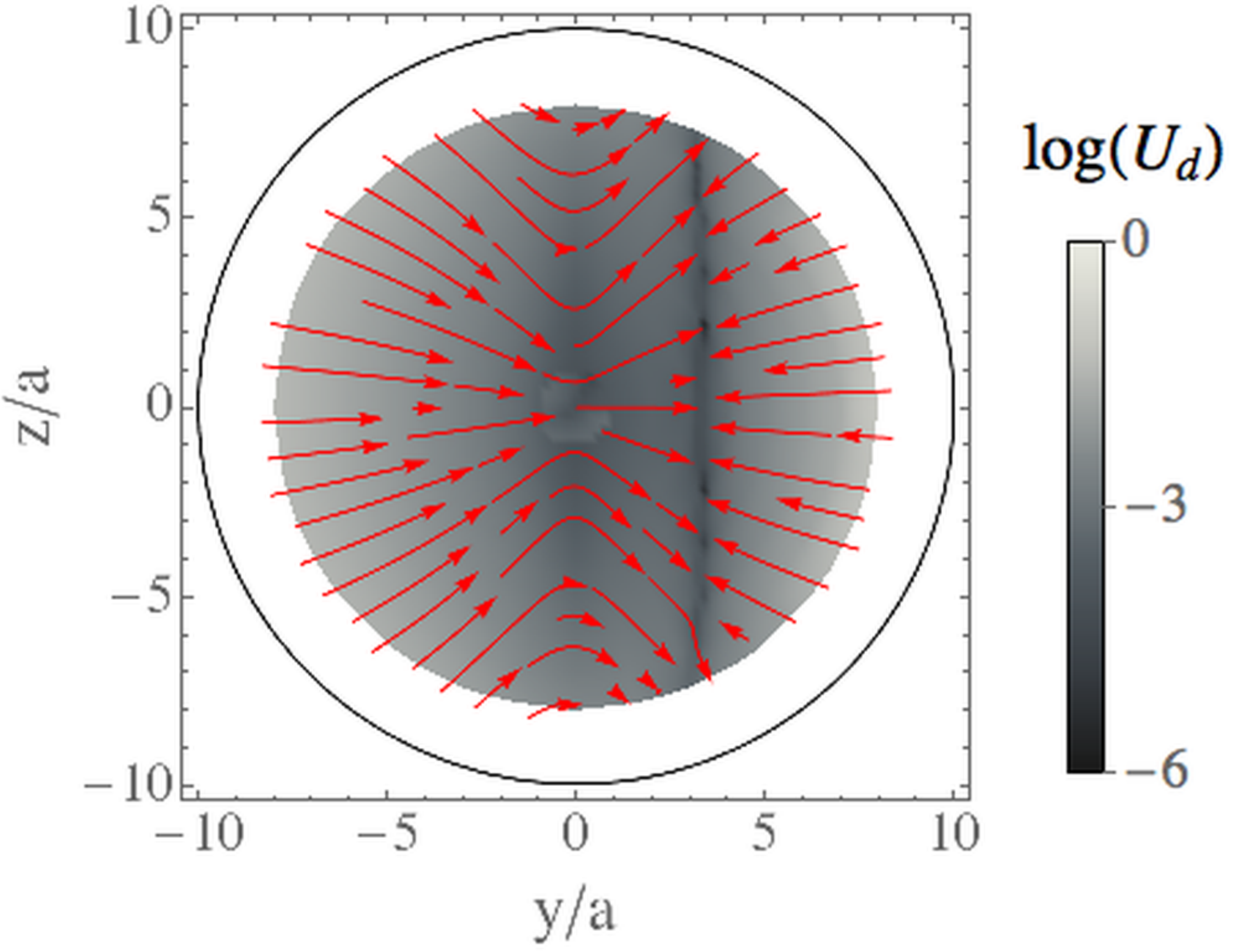} \put(-1,70){(c)}\end{overpic}} \end{minipage}
    \begin{minipage}{0.490 \columnwidth} \centerline{\begin{overpic}[width=\columnwidth]{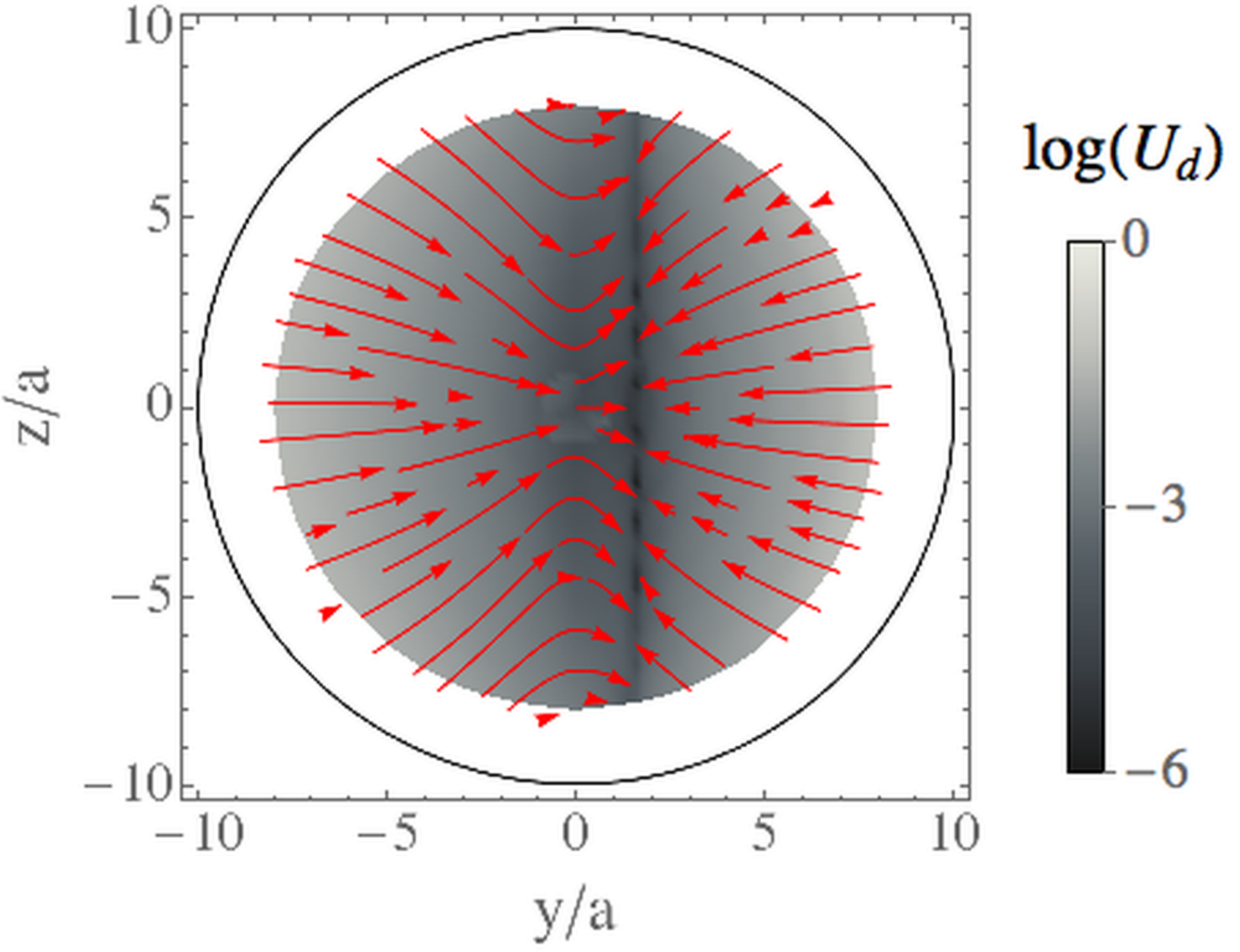} \put(-1,70){(d)}\end{overpic}} \end{minipage}
    \end{center}
    \caption{Particle trajectories projected onto the $y-z$ plane for a circular channel with radius $R/a = 10$ (black circle). Parameters are
    $\alpha = 3$, $\beta_w = 60$, and (a) $\phi_B = 0$, (b) $0.50\pi$, (c) $0.40\pi$ and (d) $0.45\pi$. \label{fig:circular}}
\end{figure}

Figure~\ref{fig:circular} shows the particle trajectories projected onto the $y-z$ plane in a circular channel with $R/a = 10$.
As for the rectangular channel, the particles move towards the closest wall for $\phi_B = 0$, while they are focused to the $y/a = 0$ plane for $\phi_B = \pi/2$. 
Interestingly if we change the direction of the magnetic field, the particles are focused to a constant $y$ plane.
The focusing region is modified with respect to the rectangular geometry because the stable angles ($\phi_p^*, \theta_p^*$) with respect to the closest wall are different. 
Since circular channels often occur in nature and are also used in microfluidic devices, focusing towards a single plane provides a particularly useful tool to manipulate particle positions.

\section{Conclusion}
We have reported a way to focus ellipsoidal magnets flowing in rectangular and circular microchannels.
A static uniform magnetic field is used to pin the particle orientation, and the particles move with translational drift velocities resulting from the image stresslets.
We derived the far-field theory predicting the particle motion in rectangular channels, and validated the accuracy of the theory by comparing the particle velocities to full numerical simulation.
The theory predicts that the particle destinations can be controlled to the walls or to the curved focusing region by changing the direction of the applied magnetic field.
Using the simulations, we also reported that the particles are focused to a single line in a circular channel. 
Our method can be used to focus and segregate magnetic particles in lab-on-a-chip devices, for example magnetically labelled rare cell species \citep{Hejazian2015}.

\section*{Acknowledgments}
This project has received funding from the European Union's Horizon 2020 research and innovation programme under grant agreement No. 665440 and under the Marie Sklodowska-Curie grant agreement No. 653284.
The authors would like to acknowledge the use of the University of Oxford Advanced Research Computing (ARC) facility in carrying out this work. http://dx.doi.org/10.5281/zenodo.22558

\bibliography{reference}

\end{document}